\documentclass[useAMS,usenatbib]{mn2e}
\usepackage{astrojournals,amsmath,graphicx,txfonts,fleqn,enumerate,color}

\newcommand {\corr}[1] {#1}
\newcommand {\correct}[1] {#1}

\newcommand*{\B}[1]   {\boldsymbol{#1}}
\newcommand*{\bxij}   {\B{x}_{i\!j}}

\newcommand*{\NH}     {N_{\!H}}
\newcommand*{\dnn}    {d_{\mathrm{nn}}}
\newcommand*{\ompar}  {\omega_{\scriptscriptstyle\parallel}}
\title[SPH without pairing instability]
      {\corr{Improving} convergence \corr{in smoothed particle hydrodynamics
          simulations} without pairing instability}

\author[Walter Dehnen \& Hossam Aly]
       {Walter Dehnen\thanks{Email: walter.dehnen@le.ac.uk,
         ha183@le.ac.uk} and Hossam Aly$^\star$\\
         Department for Physics \& Astronomy,
         University of Leicester,
         Leicester LE1 7RH
       }
\date{Accepted .
      Received ;
      }
\pagerange{\pageref{firstpage}--\pageref{lastpage}}
\pubyear{2012}
\begin{document}
\maketitle
\label{firstpage}
\begin{abstract}
  The numerical convergence of smoothed particle hydrodynamics (SPH) can be
  severely restricted by \corr{random force errors induced by} particle
  disorder, especially in shear flows\corr{, which are} ubiquitous in
  astrophysics. The increase in the number $\NH$ of neighbours when switching
  to more extended smoothing kernels \emph{at fixed resolution} (using an
  appropriate definition for the SPH resolution scale) is insufficient to
  combat these errors. Consequently, trading resolution for better convergence
  is necessary, but for traditional smoothing kernels this option is limited
  by the pairing (or clumping) instability. Therefore, we investigate the
  suitability of the Wendland functions as smoothing kernels and compare them
  with the traditional B-splines. Linear stability analysis in three
  dimensions and test simulations demonstrate that the Wendland kernels avoid
  the pairing instability for \emph{all} $\NH$, despite having vanishing
  derivative at the origin (disproving traditional ideas about the origin of
  this instability; instead, we uncover a relation with the kernel Fourier
  transform and give an explanation in terms of the SPH density
  estimator). The Wendland kernels are computationally more convenient than
  the higher-order B-splines, allowing large $\NH$ and hence better numerical
  convergence (note that computational costs rise sub-linear with $\NH$). Our
  analysis also shows that at low $\NH$ the quartic spline kernel with
  $\NH\approx60$ obtains much better convergence then the standard cubic
  spline.
\end{abstract}
\section{Introduction}
Smoothed particle hydrodynamics (SPH) is a particle-based numerical me\-thod,
pioneered by \cite{GingoldMonaghan1977} and \cite{Lucy1977}, for solving the
equations of hydrodynamics (recent reviews include \citealt{Monaghan2005,
  Monaghan2012}; \citealt{Rosswog2009}; \citealt{Springel2010:review};
\citealt{Price2012}). In SPH, the particles trace the flow and serve as
interpolation points for their neighbours. This Lagrangian nature of SPH makes
the method particularly useful for astrophysics, where typically open
boundaries apply, though it becomes increasingly popular also in engineering
\citep[e.g.][]{Monaghan2012}.

The core of SPH is the density estimator: the fluid density is
\emph{estimated} from the masses $m_i$ and positions $\B{x}_i$ of the
particles via (the symbol $\,\hat{\cdot}\,$ denotes an SPH \emph{estimate})
\begin{equation} \label{eq:rho}
  \label{eq:est:rho} \textstyle
  \rho(\B{x}_i) \approx \hat{\rho}_i \equiv 
  \sum_j m_j\,W(\B{x}_i-\B{x}_{\!j},h_i),
\end{equation}
where $W(\B{x},h)$ is the \emph{smoothing kernel} and $h_i$ the
\emph{smoothing scale}, which is adapted for each particle such that $h_i^\nu
\hat{\rho}_i=$constant (with $\nu$ the number of spatial dimensions). Similar
estimates for the value of any field can be obtained, enabling discretisation
of the fluid equations. Instead, in \emph{conservative} SPH, the equations of
motion for the particles are derived, following \cite{NelsonPapaloizou1994},
via a variational principle from the discretised Lagrangian
\begin{equation} \label{eq:L}
  \textstyle
  \mathcal{L}=\sum_i m_i 
  \left[\tfrac{1}{2}\dot{\B{x}}_i^2
    - u(\hat{\rho}_i,s_i)\right]
\end{equation}
\citep{MonaghanPrice2001}. Here, $u(\rho,s$) is the internal energy as
function of density and entropy $s$ (and possibly other gas properties), the
precise functional form of which depends on the assumed equation of state. The
Euler-Lagrange equations then yield
\begin{equation} \label{eq:hydro}
  \ddot{\B{x}}_i = \frac{1}{m_i}\frac{\partial\mathcal{L}}{\partial\B{x}_i} =
  \sum_j m_{\!j}
  \left[
    \frac{\hat{P}_i}   {\Omega_i    \hat{\rho}_i^2   }
    \B{\nabla}_{\!i}W(\bxij,h_i) +
    \frac{\hat{P}_{\!j}}{\Omega_{\!j}\hat{\rho}_{\!j}^2}
    \B{\nabla}_{\!i}W(\bxij,h_{\!j})
  \right],
\end{equation}
where $\bxij=\B{x}_i-\B{x}_{\!j}$ and $\hat{P}_i=\hat{\rho}_i^2\,\partial
u/\partial\hat{\rho}_i$, while the factors
\begin{equation} \label{eq:Omega}
  \Omega_i = 
  \frac{1}{\nu h_i^\nu\hat{\rho}_i}
  \frac{\partial(h_i^\nu\hat{\rho}_i)}{\partial \ln h_i}
  \simeq 1
\end{equation}
(\citealt{SpringelHernquist2002}; \citealt{Monaghan2002}) arise from the
adaption of $h_i$ (\citeauthor{NelsonPapaloizou1994}) such that
$h_i^\nu\hat{\rho}_i=$constant.

Equation (\ref{eq:hydro}) is a discretisation of
$\rho\ddot{\B{x}}=-\B{\nabla}\!P$, and, because of its derivation from a
variational principle, conserves mass, linear and angular momentum, energy,
entropy, and (approximately) circularity. However, its derivation from the
Lagrangian is only valid \corr{if all fluid variables are smoothly variable}.
To ensure \corr{this}, in particular \corr{for} velocity and entropy, artificial
dissipation terms have to be added to $\ddot{\B{x}}_i$ and $\dot{u}_i$. Recent
progress in restricting such dissipation to regions of compressive flow
\citep{CullenDehnen2010,ReadHayfield2012} have greatly improved the ability to
model contact discontinuities and their instabilities as well as near-inviscid
flows.

SPH is \emph{not} a Monte-Carlo method, since the particles are not randomly
distributed, but typically follow a semi-regular glass-like
distribution. Therefore, the density (and pressure) error is much smaller than
the ${\gtrsim\,}15\%$ expected from Poisson noise for ${\sim\,}40$ neighbours
and SPH obtains $\mathcal{O}(h^2)$ convergence. However, some level of particle
disorder cannot be prevented, in particular in shearing flows (as in
turbulence), where the particles are constantly re-arranged (even in the absence
of any forces), but also after a shock, where an initially isotropic particle
distribution is squashed along one direction to become anisotropic. In such
situations, the SPH force (\ref{eq:hydro}) in addition to the pressure gradient
contains a \corr{random} `\corr{E$_0$} error'
\citep*{ReadHayfieldAgertz2010}\footnote{\corr{Strictly speaking, the `E$_0$
    error' term of \citeauthor{ReadHayfieldAgertz2010} is only the dominant
    contribution to the force errors induced by particle discreteness.}},
and SPH converges more slowly than $\mathcal{O}(h^2)$. Since shocks and shear
flows are common in star- and galaxy-formation, the `\corr{E$_0$} errors' may
easily dominate the overall performance of astrophysical simulations.

One can dodge the `\corr{E$_0$} error' by using other discretisations of
$\rho\,\ddot{\B{x}}=-\B{\nabla}\!P$ \citep{Morris1996,Abel2011}. However, such
approaches unavoidably abandon momentum conservation and hence fail in practice,
in particular, for strong shocks \citep{Morris1996}. Furthermore, with such
modifications SPH no longer maintains particle order, which it otherwise
automatically achieves. Thus, the `\corr{E$_0$} error' is SPH's attempt to
resurrect particle order \citep{Price2012} and prevent shot noise from affecting
the density and pressure estimates.

Another possibility to reduce the `\corr{E$_0$} error' is to subtract an average
pressure from each particle's $\hat{P}_i$ in equation (\ref{eq:hydro}).
Effectively, this amounts to adding a negative pressure term, which can cause
the tensile instability (see \S\ref{sec:stable:cont}). Moreover, this trick is
only useful in situations with little pressure variations, perhaps\ in
simulations of near-incompressible flows \citep[e.g.][]{Monaghan2011}.

The only remaining option for reducing the `\corr{E$_0$} error' appears an
increase of the number $\NH$ of particles contributing to the density and force
estimates (contrary to naive expectation, the computational costs grow
sub-linear with $\NH$). The traditional way to try to do this is by switching to
a smoother and more extended kernel, enabling larger $\NH$ at the same smoothing
scale $h$ \citep[e.g.][]{Price2012}. However, the degree to which this approach
can reduce the `\corr{E$_0$} errors' is limited and often insufficient, even
with an infinitely extended kernel, such as the Gaussian. Therefore, one must
also consider `stretching' the smoothing kernel by increasing $h$. This
inevitably reduces the resolution, but that is still much better than obtaining
erroneous results. Of course, the best balance between reducing \corr{the `E$_0$
  error'} and resolution should be guided by results for relevant test problems
and by convergence studies.

Unfortunately, at large $\NH$ the standard SPH smoothing kernels become unstable
to the pairing (or clumping) instability (a cousin of the tensile instability),
when particles form close pairs reducing the effective neighbour number. The
pairing instability (first mentioned by \citealt{SchuesslerSchmitt1981}) has
traditionally been attributed to the diminution of the repulsive force between
close neighbours approaching each other (\citeauthor{SchuesslerSchmitt1981},
\citealt*{ThomasCouchman1992}, \citealt{Herant1994},
\citealt*{SwegleHicksAttaway1995}, \citealt{Springel2010:review},
\citealt{Price2012}). Such a diminishing near-neighbour force occurs for all
kernels with an inflection point, a necessary property of continuously
differentiable kernels. Kernels without that property have been proposed and
shown to be more stable (e.g.\ \citeauthor{ReadHayfieldAgertz2010}). However, we
provide demonstrably stable kernels with inflection point, disproving these
ideas.

Instead, our linear stability analysis in Section~\ref{sec:linear} shows that
non-negativity of the kernel Fourier transform is a necessary condition for
stability against pairing. Based on this insight we propose in
Section~\ref{sec:smooth} kernel functions, which we demonstrate in
Section~\ref{sec:test} to be indeed stable against pairing for all neighbour
numbers $\NH$, and which possess all other desirable properties. We also present
some further test simulations in Section~\ref{sec:test}, before we discuss and
summarise our findings in Sections~\ref{sec:disc} and \ref{sec:conc},
respectively.

\begin{table*}
  \begin{minipage}{160mm}
    \begin{center}
    \begin{tabular}{ll@{${\,=\,}$}lc@{\;}c@{\;}cc@{\;}c@{\;}cc@{\;\;}c@{\;\;}c}
      \multicolumn{1}{c}{kernel name} &
      \multicolumn{2}{c}{kernel function} &
      \multicolumn{3}{c}{$C$} &
      \multicolumn{3}{c}{$\sigma^2/H^2$} &
      \multicolumn{3}{c}{$H/h$}
      \\
      \multicolumn{3}{c}{} &
      $\nu=1$ & $\nu=2$ & $\nu=3$ &
      $\nu=1$ & $\nu=2$ & $\nu=3$ &
      $\nu=1$ & $\nu=2$ & $\nu=3$ \\[-1ex]
      \hline
      cubic spline & $b_4$ & $(1-r)_+^3-4(\tfrac{1}{2}-r)_+^3$ & 
      $\frac{8}{3}$ & $\frac{80}{7\pi}$ & $\frac{16}{\pi}$ &
      $\frac{1}{12}$ & $\frac{31}{392}$ & $\frac{3}{40}$ &
      1.732051 & 1.778002 & 1.825742
      \\[1ex]
      quartic spline & $b_5$ &
      $(1-r)_+^4-5(\tfrac{3}{5}-r)_+^4+10(\tfrac{1}{5}-r)_+^4$ &
      $\frac{5^5}{768}$ & $\frac{5^63}{2398\pi}$ & $\frac{5^6}{512\pi}$ &
      $\frac{1}{15}$ & $\frac{9759}{152600}$ & $\frac{23}{375}$ &
      1.936492 & 1.977173 & 2.018932
      \\[1ex]
      quintic spline & $b_6$ &
      $(1-r)_+^5-6(\tfrac{2}{3}-r)_+^5+15(\tfrac{1}{3}-r)_+^5$ &
      $\frac{3^5}{40}$ & $\frac{3^77}{478\pi}$ & $\frac{3^7}{40\pi}$ &
      $\frac{1}{18}$ & $\frac{2771}{51624}$ & $\frac{7}{135}$ &
      2.121321 & 2.158131 & 2.195775
      \\
      \hline \multicolumn{8}{c}{} \\[-2.5ex]
      Wendland $C^2$, $\nu=1$ & $\psi_{2,1}$ & $(1-r)_+^3(1+3r)$ &
      $\frac{5}{4}$ & --- & --- &
      $\frac{2}{21}$ & --- & --- &
      1.620185 & --- & ---
      \\[1ex]
      Wendland $C^4$, $\nu=1$ & $\psi_{3,2}$ & $(1-r)_+^5(1+5r+8r^2)$ &
      $\frac{3}{2}$ & --- & --- &
      $\frac{1}{15}$ & --- & --- &
      1.936492 & --- & ---
      \\[1ex]
      Wendland $C^6$, $\nu=1$ & $\psi_{4,3}$ & $(1-r)_+^7(1+7r+19r^2+21r^3)$ &
      $\frac{55}{32}$ & --- & --- &
      $\frac{2}{39}$ & --- & --- &
      2.207940 & --- & ---
      \\
      \hline \multicolumn{8}{c}{} \\[-2.5ex]
      Wendland $C^2$, $\nu=2,3$ & $\psi_{3,1}$ & $(1-r)_+^4(1+4r)$ &
      --- & $\frac{7}{\pi}$ & $\frac{21}{2\pi}$ &
      --- & $\frac{5}{72}$ & $\frac{1}{15}$ &
      --- & 1.897367 & 1.936492
      \\[1ex]
      Wendland $C^4$, $\nu=2,3$ & $\psi_{4,2}$ &
      $(1-r)_+^6(1+6r+\frac{35}{3}r^2)$ &
      --- & $\frac{9}{\pi}$ & $\frac{495}{32\pi}$ &
      --- & $\frac{7}{132}$ & $\frac{2}{39}$ &
      --- & 2.171239 & 2.207940
      \\[1ex]
      Wendland $C^6$, $\nu=2,3$ & $\psi_{5,3}$ & 
            $(1-r)_+^8(1+8r+25r^2+32r^3)$ &
      --- & $\frac{78}{7\pi}$ & $\frac{1365}{64\pi}$ &
      --- & $\frac{3}{70}$ & $\frac{1}{24}$ &
      --- & 2.415230 & 2.449490
    \end{tabular}
    \caption{\label{tab:kernel} Functional forms and various quantities for the
      B-splines (equation~\ref{eq:B-spline}) and Wendland functions
      (equation~\ref{eq:Wendland}) in $\nu=1$-3 spatial
      dimensions. $(\cdot)_+\equiv\max\{0,\cdot\}$. $C$ is the normalisation
      constant, $\sigma$ the standard deviation (equation~\ref{eq:sigma}), and
      $h=2\sigma$ the smoothing scale. Note that the Wendland functions of given
      differentiability are identical for $\nu=2$ and $\nu=3$ but differ from
      those for $\nu=1$.  $\psi_{2,1}$ (the $C^2$ Wendland function in 1D) has
      already been used in the second SPH paper ever \citep{Lucy1977}, but for
      3D simulations, when it is not a Wendland function.}
    \end{center}
  \end{minipage}
\end{table*}
%
\section{Smoothing matters} \label{sec:smooth}
\subsection{Smoothing scale}\label{sec:smooth:scale}
SPH smoothing kernels are usually isotropic and can be written as
\begin{equation} \label{eq:W:h}
  W(\B{x},h) = h^{-\nu}\,\tilde{w}(|\B{x}|/h)
\end{equation}
with a dimensionless function $\tilde{w}(r)$, which specifies the functional
form and satisfies the normalisation $1=\!\int\mathrm{d}^\nu\!\B{x}
\,\tilde{w}(|\B{x}|)$. The re-scaling $h\to\alpha h$ and
$\tilde{w}(r)\to\alpha^\nu\tilde{w}(\alpha r)$ with $\alpha>0$ leaves the
functional form of $W(\B{x})$ unchanged but alters the meaning of $h$. In order
to avoid this ambiguity, a definition of the smoothing scale in terms of the
kernel, i.e.\ via a functional $h=h[W(\B{x})]$, must be specified.

In this study we use two scales, the smoothing scale $h$, defined below, and the
\emph{kernel-support radius} $H$, the largest $|\B{x}|$ for which
$W(\B{x})>0$. For computational efficiency, smoothing kernels used in practice
have compact support and hence finite $H$. For such kernels
\begin{equation} \label{eq:W}
  W(\B{x},h) = H^{-\nu}\,w(|\B{x}|/H),
\end{equation}
where $w(r)=0$ for $r\ge1$ and $w(r)>0$ for $r<1$. $H$ is related to the average
number $\NH$ of neighbours within the smoothing sphere by
\begin{equation} \label{eq:NH}
  \NH = V_{\!\nu}\,H_i^\nu(\hat{\rho}_i/m_i)
\end{equation}
with $V_\nu$ the volume of the unit sphere. $H$ and $\NH$ are useful quantities
in terms of kernel computation and neighbour search, but not good measures for
the smoothing scale $h$. Unfortunately, there is some confusion in the SPH
literature between $H$ and $h$, either being denoted by `$h$' and referred to as
`smoothing length'. Moreover, an appropriate definition of $h$ in terms of the
smoothing kernel is lacking. Possible definitions include the kernel standard
deviation
\begin{equation} \label{eq:sigma}
  \textstyle
  \sigma^2
  = \nu^{-1} \int\mathrm{d}^\nu\!\B{x}\,\B{x}^2\,W(\B{x},h),
\end{equation}
the radius of the inflection point (maximum of $|\B{\nabla}W|$), or the ratio
$W/|\B{\nabla}W|$ at the inflection point. For the Gaussian kernel
\begin{equation} \label{eq:gaussian}
  W(\B{x}) = \mathcal{N}(0,\sigma^2) \equiv
  \frac{1}{(2\pi\sigma^2)^{\nu/2}}\,
  \exp\left(-\frac{\B{x}^2}{2\sigma^2}\right)
\end{equation}
all these give the same result independent of dimensionality, but not for
other kernels (\correct{`triangular' kernels have no inflection
  point}). Because the standard deviation (\ref{eq:sigma}) is directly related
to the numerical resolution of sound waves (\S\ref{sec:stable:long}), we set
\begin{equation} \label{eq:h}
  h=2\sigma.
\end{equation}

In practice (and in the remainder of our paper), the neighbour number $\NH$ is
often used as a convenient parameter, even though it holds little meaning by
itself. A more meaningful quantity in terms of resolution is the average number
$N_{\!h}$ of particles within distance $h$, given by $N_{\!h}\equiv(h/H)^\nu\NH$
for kernels with compact support, or the ratio \corr{$h(\hat{\rho}/m)^{1/\nu}$
  between $h$ and the average particle separation}.

\subsection{Smoothing kernels} \label{sec:smooth:kern}
After these definitions, let us list the desirable properties of the smoothing
kernel \citep[cf.][]{FulkQuinn1996,Price2012}.
\begin{enumerate}[(i)]
  \label{list:kernel}
  \item \label{enum:1} equation (\ref{eq:rho}) obtains an accurate density
    estimate;
  \item \label{enum:2} $W(\B{x},h)$ is twice continuously
    differentiable;
  \item \label{enum:4} SPH is stable against pairing at the desired
    $\NH$;
  \item \label{enum:5} $W(\B{x},h)$ and $\B{\nabla}W(\B{x},h)$
    are computationally inexpensive.
\end{enumerate}
Here, condition (\ref{enum:1}) implies that $W(\B{x},h)\to\delta(\B{x})$ as
$h\to0$ but also that $W(\B{x},h)\ge0$ is monotonically declining with
$|\B{x}|$; condition (\ref{enum:2}) guarantees smooth forces, but also implies
$\B{\nabla}W(0)=0$.

\subsubsection{B-splines} \label{sec:B-splines}
The most used SPH kernel functions are the \cite{Schoenberg1946} B-spline
functions, generated as 1D Fourier transforms\footnote{By this definition they
  are the $n$-fold convolution \corr{(in one dimension)} of $b_1(r)$ with itself
  (modulo a scaling), and hence are identical to the
  \cite{Irwin1927}-\cite{Hall1927} probability density for the sum $r$ of $n$
  independent random variables, each uniformly distributed between $-1/n$ and
  $1/n$.\label{foot:irwin:hall}} \citep{MonaghanLattanzio1985}
\begin{equation} \label{eq:B-spline}
  w(r) = C\,b_n(r),
  \quad b_n(r)\equiv \frac{1}{2\pi}\int_{-\infty}^\infty
  \left(\frac{\sin k/n}{k/n}\right)^n\cos kr\;\mathrm{d}k,
\end{equation}
with normalisation constant $C$. These kernels consist of $\lceil n/2\rceil$
piece-wise polynomials of degree $n-1$ (see Table~\ref{tab:kernel}) and are
$n-2$ times continuously differentiable. Thus, the cubic spline ($n=4$) is the
first useful, but the quartic and quintic have also been used. For large $n$,
the B-splines approach the Gaussian: $b_n \to \mathcal{N}(0,H^2\!/3n)$ (this
follows from footnote~\ref{foot:irwin:hall} and the central limit theorem).

Following \citeauthor{MonaghanLattanzio1985}, $\tilde{h}\equiv2H/n$ is
conventionally used as smoothing scale for the B-splines independent of
$\nu$. This is motivated by their original purpose to interpolate equidistant
one-dimensional data with spacing $\tilde{h}$, but cannot be expressed via a
functional $\tilde{h}=\tilde{h}[W(\B{x})]$. Moreover, the resulting ratios
between $\tilde{h}$ for the $b_n$ do not match any of the definitions discussed
above\footnote{Fig.~2 of \cite{Price2012} seems to suggest that with this
  scaling the B-splines approach the Gaussian with
  $\sigma=\tilde{h}/\!\!\sqrt{2}$. However, this is just a coincidence for $n=6$
  (quintic spline) since $\sigma=\!\!\sqrt{n/12}\,\tilde{h}$ for the B-splines
  in 1D.}.

Instead, we use the more appropriate $h=2\sigma$ also for the B-spline kernels,
giving $H\approx1.826\,h$ for the cubic spline in 3D, close to the conventional
$H=2\tilde{h}$ (see Table~\ref{tab:kernel}).

\subsubsection{`Triangular' kernels} \label{sec:kern:triang}
At low order $n$ the B-splines are only stable against pairing for
modest values of $\NH$ (we will be more precise in Section~\ref{sec:linear}),
while at higher $n$ they are computationally increasingly complex.

Therefore, alternative kernel functions which are stable for large $\NH$ are
desirable. As the pairing instability has traditionally been associated with the
presence of an inflection point (minimum of $w'$), functions $w(r)$ without
inflection point have been proposed. These have a triangular shape at $r\sim0$
and necessarily violate point (ii) of our list, but avoid the pairing\correct{}
instability\footnote{\cite{ThomasCouchman1992} proposed such kernels only for
  the force equation (\ref{eq:hydro}), but to keep a smooth kernel for the
  density estimate. However, such an approach cannot be derived from a
  Lagrangian and hence necessarily violates energy and/or entropy conservation
  \citep{Price2012}.}. For comparison we consider one of them, the `HOCT4'
kernel of \cite{ReadHayfieldAgertz2010}.

\subsubsection{Wendland functions} \label{sec:kern:wend}
The linear stability analysis of the SPH algorithm, presented in the next
Section, shows that a necessary condition for stability against pairing is the
non-negativity of the multi-dimensional Fourier transform of the kernel. The
Gaussian has non-negative Fourier transform for any dimensionality and hence
would give an ideal kernel were it not for its infinite support and
computational costs.

\begin{figure}
  \begin{center}
    \resizebox{75mm}{!}{\includegraphics{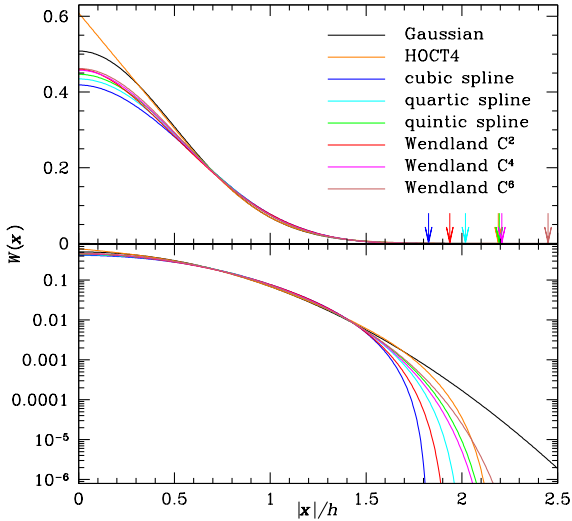}}
  \end{center}
  \vspace*{-3ex}
 \caption{
    \label{fig:kernel}
    Kernels of Table~\ref{tab:kernel}, the Gaussian, and the HOCT4 kernel of
    Read et al.~(2010) scaled to a common resolution scale of $h=2\sigma$ for 3D
    (\emph{top}: linear plot, arrows indicating $|\B{x}|= H$; \emph{bottom}:
    logarithmic plot).  }
\end{figure}

Therefore, we look for kernel functions of compact support which have
non-negative Fourier transform in $\nu$ dimensions and are low-order
polynomials\footnote{Polynomials in $r^2$ would avoid the computation of a
  square root. However, it appears that such functions cannot possibly have
  non-negative Fourier transform (H.~Wendland, private communication).} in $r$.
This is precisely the defining property of the \cite{Wendland1995} functions,
which are given by
\begin{equation} \label{eq:Wendland}
  w(r)=C\,\psi_{\ell k}(r),
  \quad
  \psi_{\ell k}(r) \equiv \mathcal{I}^k\,(1-r)_+^\ell
\end{equation}
with $(\cdot)_+^{}\equiv\max\{0,\cdot\}$ and the linear operator
\begin{equation}
  \textstyle
  \mathcal{I}[f](r) \equiv \int_r^\infty sf(s)\,\mathrm{d}s.
\end{equation}
In $\nu$ spatial dimensions, the functions $\psi_{\ell k}(|\B{x}|)$ with $\ell=
k+1+\lfloor\nu/2\rfloor$ have positive Fourier transform and are $2k$ times
continuously differentiable. In fact, they are the unique polynomials in
$|\B{x}|$ of minimal degree with these properties
\citep{Wendland1995,Wendland2005}. For large $k$, they approach the Gaussian,
which is the only non-trivial eigenfunction of the operator $\mathcal{I}$. We
list the first few Wendland functions for one, two, and three dimensions in
Table~\ref{tab:kernel}, and plot them for $\nu=3$ in Fig.~\ref{fig:kernel}.

\subsection{Kernel comparison} \label{sec:kern:comp}
Fig.~\ref{fig:kernel} plots the kernel functions $w(r)$ of
Table~\ref{tab:kernel}, the Gaussian, and the HOCT4 kernel, all scaled to the
same $h=2\sigma$ for $\nu=3$. Amongst the various scalings (ratios for $h/H$)
discussed in \S\ref{sec:smooth:scale} above, this gives by far the best match
between the kernels.  The B-splines and Wendland functions approach the Gaussian
with increasing order. The most obvious difference between them in this scaling
is their central value. The B-splines, in particular of lower order, put less
emphasis on small $r$ than the Wendland functions or the Gaussian.

Obviously, the HOCT4 kernel, which has no inflection point, differs
significantly from all the others and puts even more emphasis on the centre than
the Gaussian (for this kernel $\sigma\approx0.228343H$).

\subsection{Kernel Fourier transforms} \label{sec:kern:comp:four}
For spherical kernels of the form (\ref{eq:W}), their Fourier transform only
depends on the product $H|\B{k}|$, i.e.\ $\widehat{W}(\B{k}) =
\widehat{w}(H|\B{k}|)$. In 3D ($\mathcal{F}_\nu$ denotes the Fourier transform
in $\nu$ dimensions)
\begin{equation} \label{eq:w:kappa}
  \textstyle
  \widehat{w}(\kappa) = \mathcal{F}_3\left[w(r)\right](\kappa) =
  4\pi \kappa^{-1} \int_0^\infty \sin(\kappa r)\,w(r)\,r\,\mathrm{d}r
\end{equation}
which is an even function and (up to a normalisation constant) equals
$-\kappa^{-1}\mathrm{d}\mathcal{F}_1\big[w\big]/\mathrm{d}\kappa$. For the
B-splines, which are defined via their 1D Fourier transform in equation
(\ref{eq:B-spline}), this gives immediately
\begin{equation}
  \label{eq:w:kappa:B}
  \textstyle
  \mathcal{F}_3\left[b_n(r)\right](\kappa) = 3 \left(\frac{\textstyle
    n}{\textstyle\kappa}\right)^{n+2}
  \sin^n\!\frac{\textstyle\kappa}{\textstyle n} \left(1-
  \frac{\textstyle\kappa}{\textstyle n}\cot\frac{\textstyle\kappa}{\textstyle
    n}\right)
\end{equation}
(which includes the normalisation constant), while for the 3D Wendland kernels
\begin{equation}
  \label{eq:w:kappa:W}
  \mathcal{F}_3\left[\psi_{\ell k}(r)\right](\kappa) = 
  \left(-\frac{1}{\kappa}\frac{\mathrm{d}}{\mathrm{d}\kappa}\right)^{k+1}
  \mathcal{F}_1\left[(1-r)^\ell_+\right](\kappa)
\end{equation}
(we abstain from giving individual functional forms).
\begin{figure}
  \begin{center}
    \resizebox{80mm}{!}{\includegraphics{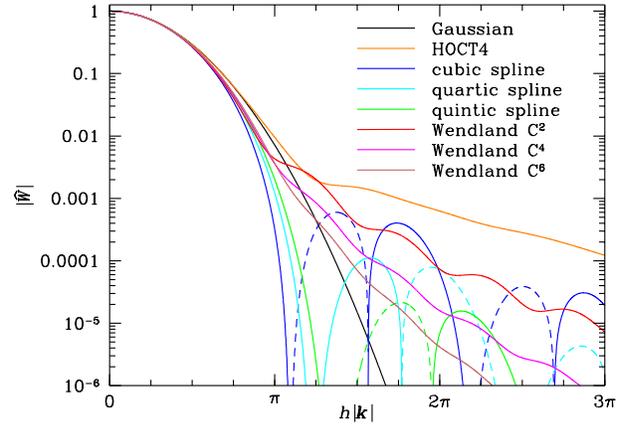}}
  \end{center}
  \vspace*{-3ex}
  \caption{
    \label{fig:fk}
    Fourier transforms $\widehat{W}(\B{k})$ for the Gaussian, the HOCT4 and the
    kernels of Table~\ref{tab:kernel} scaled to the same common scale
    $h=2\sigma$. Negative values are plotted with broken curves.  }
\end{figure}

All these are plotted in Fig.~\ref{fig:fk} after scaling them to a common
$h=2\sigma$. Notably, all the B-spline kernels obtain $\widehat{W}<0$ and
oscillate about zero for large $\B{k}$ (which can also be verified directly from
equation~\ref{eq:w:kappa:B}), whereas the Wendland kernels have
$\widehat{W}(\B{k})>0$ at all $\B{k}$, as does the HOCT4 kernel. As
non-negativity of the Fourier transform is necessary (but not sufficient) for
stability against pairing at large $\NH$ (see \S\ref{sec:stable:cont}), in 3D
the B-splines (of any order) fall prey to this instability for sufficiently
large $\NH$, while, based solely on their Fourier transforms, the Wendland and
HOCT4 kernels may well be stable for all neighbour numbers.

At large $|\B{k}|$ (small scales), the HOCT kernel has most power, caused by its
central spike, while the other kernels have ever less small-scale power with
increasing order, becoming ever smoother and approaching the Gaussian, which has
least small-scale power.

The scaling to a common $h=2\sigma$ in Fig.~\ref{fig:fk} has the effect that the
$\widehat{W}(\B{k})$ all overlap at small wave numbers, since their Taylor
series
\begin{equation} \label{eq:w:Taylor}
  \widehat{W}(\B{k}) = 1 - \tfrac{1}{2}\sigma^2 \B{k}^2 +
  \mathcal{O}(|\B{k}|^4).
\end{equation}

\subsection{Density estimation and correction} \label{sec:kern:dens}
The SPH force (\ref{eq:hydro}) is inseparably related, owing to its derivation
via a variational principle, to the \emph{derivative} of the density
estimate. Another important role of the SPH density estimator is to obtain
accurate values for $\hat{P}_i$ in equation (\ref{eq:hydro}), and we will now
assess the performance of the various kernels in this latter respect.

\begin{figure}
  \begin{center}
    \resizebox{80mm}{!}{\includegraphics{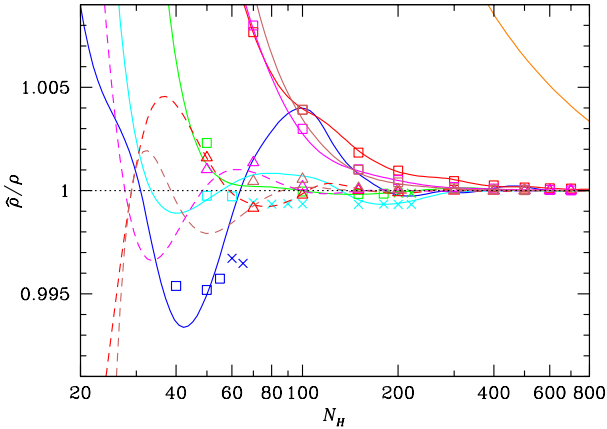}}
  \end{center}
  \vspace*{-3ex}
  \caption{
    \label{fig:rho}
    SPH density estimate (\ref{eq:rho}) obtained from particles in
    three-dimensional densest-sphere packing (\emph{solid}), glass
    (\emph{squares}), or pairing (\emph{crosses}), plotted against neighbour
    number $\NH$ for the kernels of Table~\ref{tab:kernel} and the HOCT4 kernel
    (colour coding as in Figs.~\ref{fig:kernel}\&\ref{fig:fk}). For the Wendland
    kernels, the corrected density estimate (\ref{eq:rho:corr}) is shown with
    \emph{dashed} curves and \emph{triangles} for densest-sphere packing, and a
    glass, respectively.  }
\end{figure}

In Fig.~\ref{fig:rho}, we plot the estimated density (\ref{eq:rho})
vs.\ neighbour number $\NH$ for the kernels of Table~\ref{tab:kernel} and
particles distributed in three-dimensional densest-sphere packing (solid
curves) or a glass (squares). While the standard cubic spline kernel
under-estimates the density (only values $\NH\lesssim\correct{55}$ are
accessible for this kernel owing to the pairing instability), the Wendland
kernels (and Gaussian, not shown) tend to over-estimate it.

It is worthwhile to ponder about the origin of this density
over-estimation. If the particles were randomly rather than semi-regularly
distributed, $\hat{\rho}$ obtained for an unoccupied position would be
unbiased \citep[e.g.][]{Silverman1986}, while at a particle position the self
contribution $m_iW(0,h_i)$ to $\hat{\rho}_i$ results in an over-estimate. Of
course, in SPH and in Fig.~\ref{fig:rho} particles are not randomly
distributed, but at small $\NH$ the self-contribution still induces some bias,
as evident from the over-estimation for \emph{all} kernels at very small
$\NH$.

The HOCT4 kernel of Read et al.~(2010, \emph{orange}) with its central spike
(cf.~Fig.~\ref{fig:kernel}) shows by far the worst performance. However, this is
not a peculiarity of the HOCT4 kernel, but a generic property of all kernels
without inflection point.

These considerations suggest the \emph{corrected} density estimate
\begin{equation}\label{eq:rho:corr}
  \hat{\rho}_{i,\mathrm{corr}} = \hat{\rho}_i - \epsilon\,m_i\, W(0,h_i),
\end{equation}
which is simply the original estimate (\ref{eq:rho}) with a fraction $\epsilon$
of the self-contribution subtracted. \corr{The equations of motion obtained by
  replacing $\hat{\rho}_i$ in the Lagrangian (\ref{eq:L}) with
  $\hat{\rho}_{i,\mathrm{corr}}$ are otherwise identical to equations
  (\ref{eq:hydro}) and (\ref{eq:Omega}) (note that $
  \partial(\!h_i^\nu\hat{\rho}^{}_{i,\mathrm{corr}})/\partial\ln h_i=
  \partial(\!h_i^\nu\hat{\rho}^{}_i)/\partial\ln h_i$, since
  $h_i^\nu\hat{\rho}_i$ and $h_i^\nu\hat{\rho}_{i,\mathrm{corr}}$ differ only by
  a constant), in particular the} conservation properties are unaffected.  From
the data of Fig.~\ref{fig:rho}, we find that good results are obtained by a
simple power-law
\begin{equation}\label{eq:eps:Nnb}
  \epsilon = \epsilon_{100}\, (\NH/100)^{-\alpha}
\end{equation}
with constants $\epsilon_{100}$ and $\alpha$ depending on the kernel.  We use
$(\epsilon_{100},\alpha)$ = (0.0294,\,0.977), (0.01342,\,1.579), and
(0.0116,\,2.236), respectively, for the Wendland $C^2$, $C^4$, and $C^6$ kernels
in $\nu=3$ dimensions.

The dashed curves and triangles in Fig.~\ref{fig:rho} demonstrate that this
approach obtains accurate density and hence pressure estimates.

\section{Linear stability and sound waves} \label{sec:linear}
The SPH linear stability analysis considers a plane-wave perturbation to an
equilibrium configuration, i.e.\ the positions are perturbed according to
\begin{equation} \label{eq:x:pert:}
  \B{x}_{i}\to\B{x}_i+\B{a}\,\exp\big(i[\B{k}{\cdot}\B{x}_{i}-\omega t]\big)
\end{equation}
with displacement amplitude $\B{a}$, wave vector $\B{k}$, and angular frequency
$\omega$. Equating the forces generated by the perturbation to linear order in
$\B{a}$ to the acceleration of the perturbation yields a dispersion relation of
the form
\begin{equation} \label{eq:dispersion}
  \B{a}\cdot\B{\mathsf{P}}(\B{k}) = \omega^2\B{a}.
\end{equation}
This is an eigenvalue problem for the matrix $\B{\mathsf{P}}$ with eigenvector
$\B{a}$ and eigenvalue $\omega^2$. The exact (non-SPH) dispersion relation
(with $c^2=\partial P/\partial \rho$, $P=\rho^2\partial u/\partial\rho$ at
\correct{constant} entropy)
\begin{equation} \label{eq:P:exact}
  c^2\,\B{a} {\cdot} \B{k}\,\B{k} = \omega^2\B{a}
\end{equation}
has only one non-zero eigenvalue $\omega^2= c^2\B{k}^2$ with eigenvector
$\B{a}\parallel\B{k}$, corresponding to longitudinal sound waves propagating at
speed $c$.

The actual matrix $\B{\mathsf{P}}$ in equation (\ref{eq:dispersion}) depends on
the details of the SPH algorithm. For conservative SPH with equation of motion
(\ref{eq:hydro}), \cite{Monaghan2005} gives it for $P\propto\rho^\gamma$ in one
spatial dimension. We derive it in appendix~\ref{app:linear} for a general
equation of state and any number $\nu$ of spatial dimensions:
\begin{equation} \label{eq:P}
  \B{\mathsf{P}} =
  \bar{c}^2 \B{u}^{(2)}
  +
  \frac{2\bar{P}}{\bar{\rho}}
  \left(\B{\mathsf{U}}
  -\B{u}^{(2)}
  +\!
  \left\{
  \frac{\bar{\Xi}}{2}\B{u}^{(2)}
  -
  \frac{1}{2\nu\bar{\rho}^2\bar{\Omega}^2}
  \frac{\partial\B{t}^{(2)}}{\partial\ln\!\bar{h}}
  \right\}\right),
\end{equation}
where $\B{x}^{(2)}$ is the outer product of a vector with itself, bars denote
SPH estimates for the unperturbed equilibrium,
$\B{t}=\bar{\rho}\bar{\Omega}\B{u}$, and
\begin{subequations}
  \label{eq:uUPi}
  \begin{eqnarray}
    \label{eq:u}
    \B{u}(\B{k}) &=& \frac{1}{\bar{\rho}\{\bar{\Omega}\}}
    \sum_j m \sin\B{k}{\cdot}\bar{\B{x}}_{\!j}\,\B{\nabla}
    W(\bar{\B{x}}_{\!j},\bar{h}),
    \\
    \label{eq:U}
    \B{\mathsf{U}}(\B{k}) &=& \frac{1}{\bar{\rho}\{\bar{\Omega}\}}
    \sum_j m \,\big(1-\cos\B{k}{\cdot}\bar{\B{x}}_{\!j}\big)\,
    \B{\nabla}^{(2)}W(\bar{\B{x}}_{\!j},\bar{h}),
    \\
    \label{eq:Xi}
    \bar{\Xi} &=& \frac{1}{\nu}
    \frac{\partial\ln(\bar{\rho}\bar{\Omega})}{\partial\ln\bar{h}}
    \simeq 0.
  \end{eqnarray}
\end{subequations}
Here and in the remainder of this section, curly brackets indicate terms not
present in the case of a constant $h=\bar{h}$, when our results reduce to
relations given by \cite{Morris1996} and \cite{ReadHayfieldAgertz2010}.

Since $\B{\mathsf{P}}$ is real and symmetric, its eigenvalues are real and its
eigenvectors mutually orthogonal\footnote{If in equation (\ref{eq:hydro}) one
  omits the factors $\Omega_i$ but still adapts $h_i$ to obtain
  $h_i^\nu\hat{\rho}_i=$constant, as some practitioners do, then the resulting
  dispersion relation has an asymmetric matrix $\B{\mathsf{P}}$ with potentially
  complex eigenvalues.}. The SPH dispersion relation (\ref{eq:dispersion}) can
deviate from the true relation (\ref{eq:P:exact}) in mainly two ways. First, the
longitudinal eigenvalue $\ompar^2$ (with eigenvector
$\B{a}_{\scriptscriptstyle\parallel}\parallel\B{k}$) may deviate from
$c^2\B{k}^2$ (wrong sound speed) or even be negative (pairing instability;
\citealt{Morris1996, Monaghan2000}). Second, the \correct{other two} eigenvalues
$\omega^2_{\perp1,2}$ may be significantly non-zero (transverse instability for
$\omega^2_{\perp1,2}<0$ or transverse sound waves for $\omega^2_{\perp1,2}>0$).

\begin{figure*}
  \begin{center}
    \resizebox{86.25mm}{!}{\includegraphics{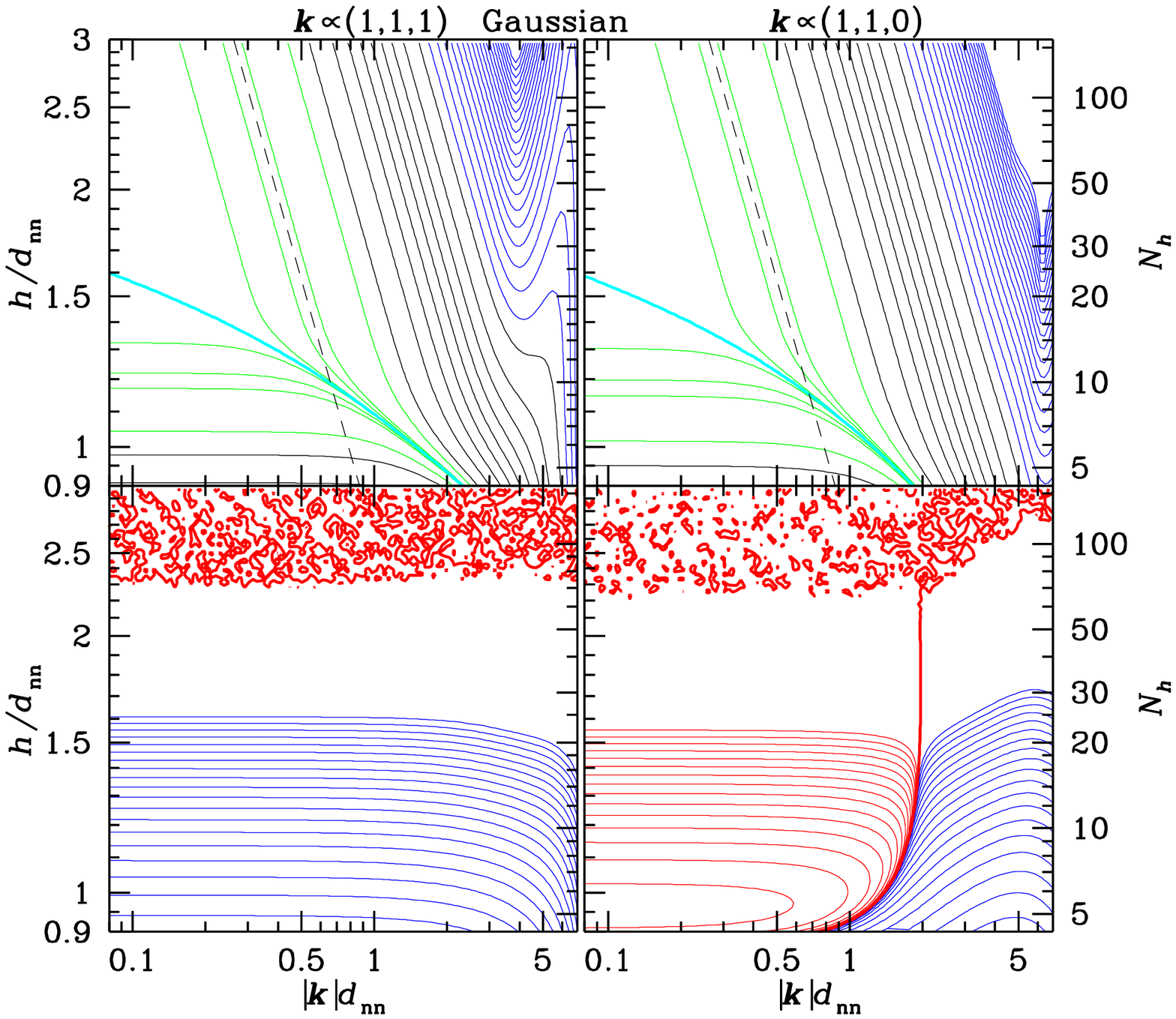}} \hfil
    \resizebox{86.25mm}{!}{\includegraphics{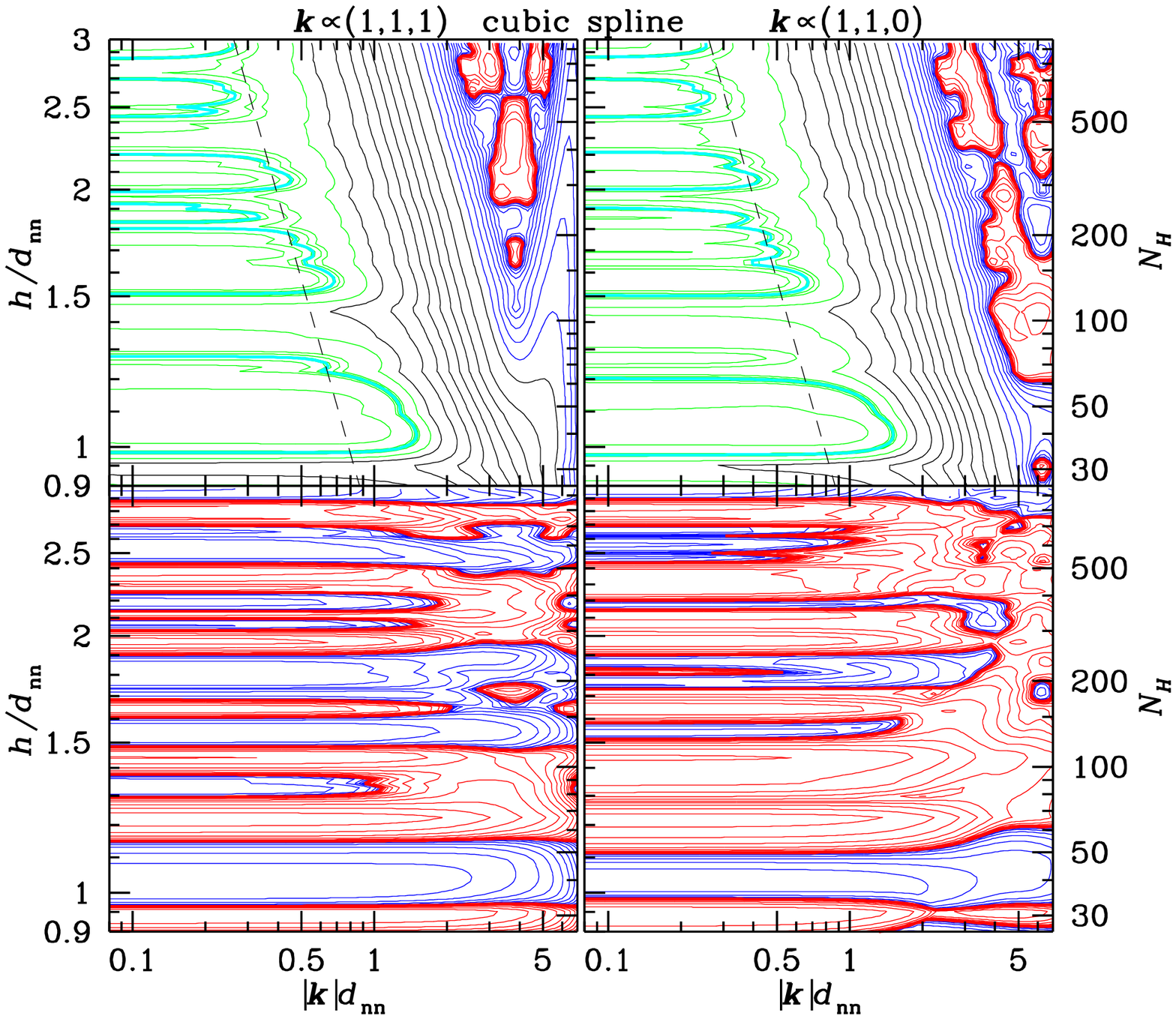}}
  \end{center}
  \vspace*{-3ex}
  \caption{
    \label{fig:stable:s3}
    Stability of conservative SPH with the Gaussian (truncated at $16\sigma$)
    and cubic spline kernel for densest-sphere packing and $P\propto\rho^{5/3}$:
    contours of $\ompar^2/c^2\B{k}^2$ (\emph{top}) and
    $\omega^2_{\perp2}/c^2\B{k}^2$ (\emph{bottom}), the longitudinal and
    smallest transverse eigenvalues of $\B{\mathsf{P}}/c^2\B{k}^2$ respectively,
    over wave number $|\B{k}|$ and smoothing scale $h=2\sigma$ (and $\NH$ or
    $N_{\!h}$) both scaled to the nearest-neighbour distance $\dnn$. The
    \emph{left} and \emph{right} sub-panels are for $\B{k}\propto(1,1,1)$
    (perpendicular to hexagonal planes), and $\B{k}\propto(1,1,0)$
    (nearest-neighbour direction), respectively (other wave vectors give similar
    results). Red contours are for $\omega^2\le0$ (implying the pairing
    instability in the \emph{top} panels) and are logarithmically spaced by
    $0.25$ dex. Blue contours are also logarithmically spaced between $10^{-6}$
    and $0.1$, black contours are linearly spaced by $0.1$, while good values
    for $\ompar^2$ are $\ompar^2/c^2\B{k}^2=0.95$, 0.99, 0.995, 0.999, 1.001
    1.005, 1.01, 1.05 (\emph{green}), and 1 (\emph{cyan}). The dashed line
    indicates a sound wave with wavelength $\lambda=8h=16\sigma$. For the
    Gaussian kernel $|\omega^2_{\perp2}|$ in the bottom panels is often smaller
    than our numerical precision.  }
\end{figure*}

\begin{figure*}
  \begin{center}
    \resizebox{86.25mm}{!}{\includegraphics{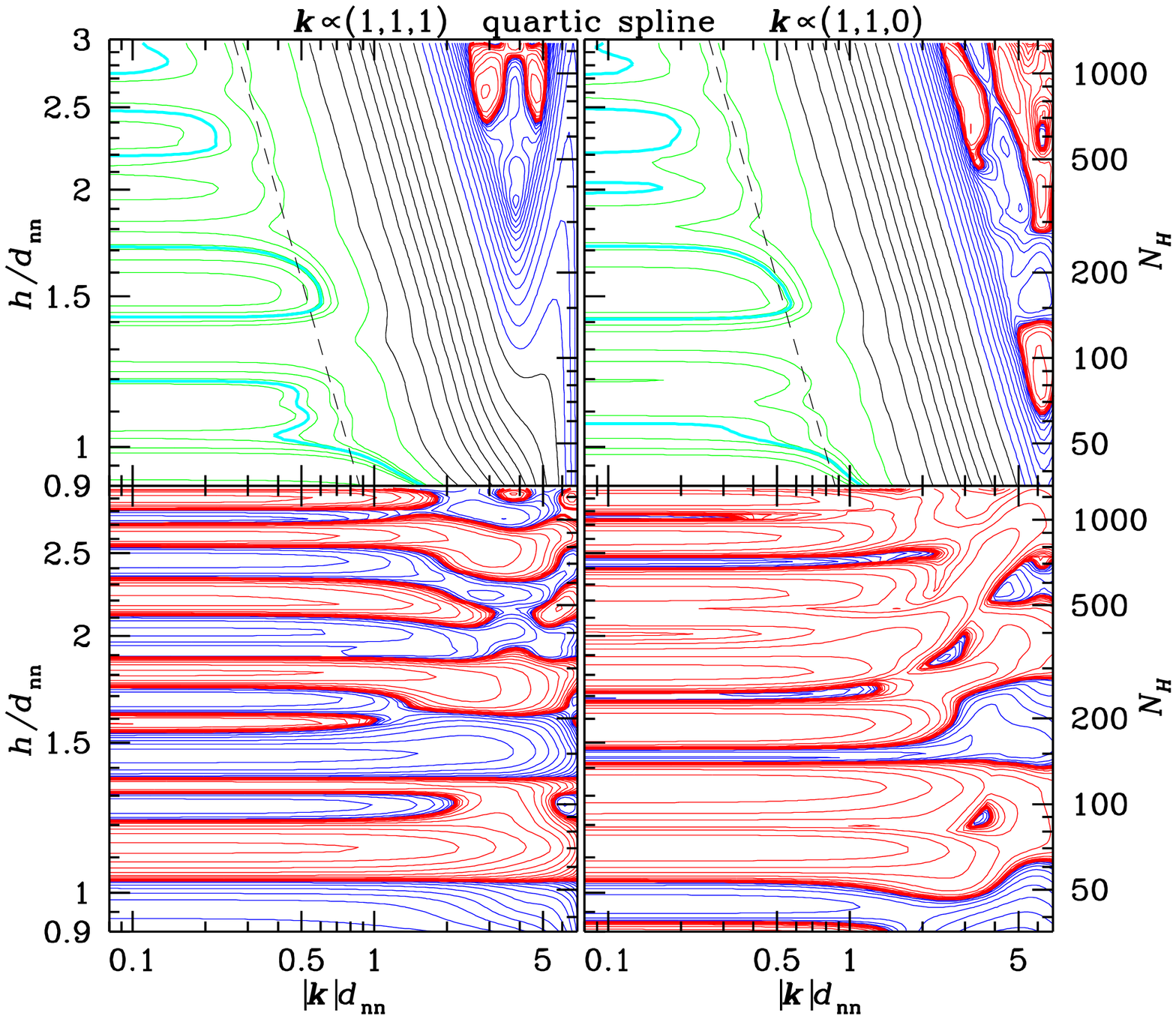}} \hfil
    \resizebox{86.25mm}{!}{\includegraphics{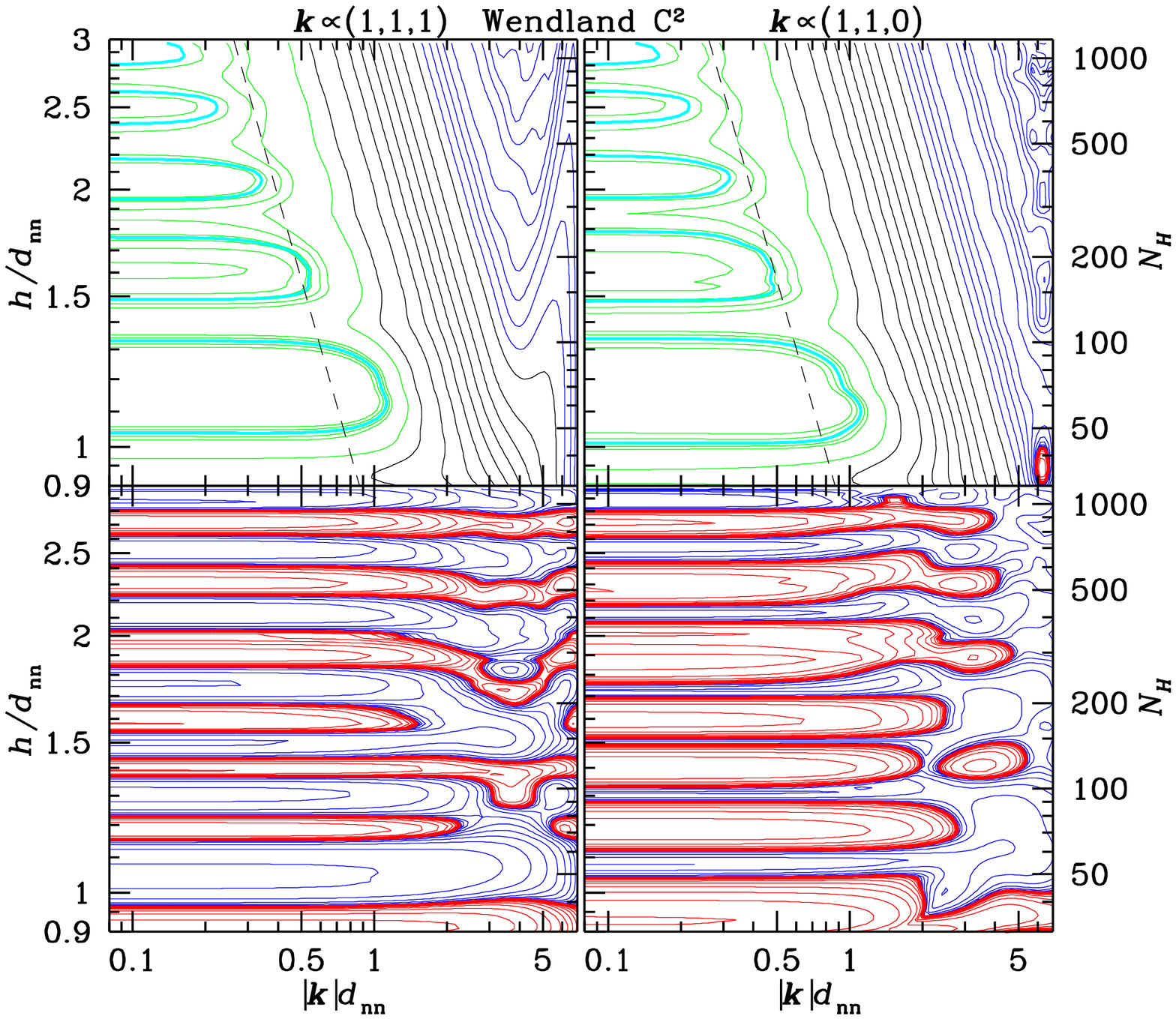}} \\[2ex]
    \resizebox{86.25mm}{!}{\includegraphics{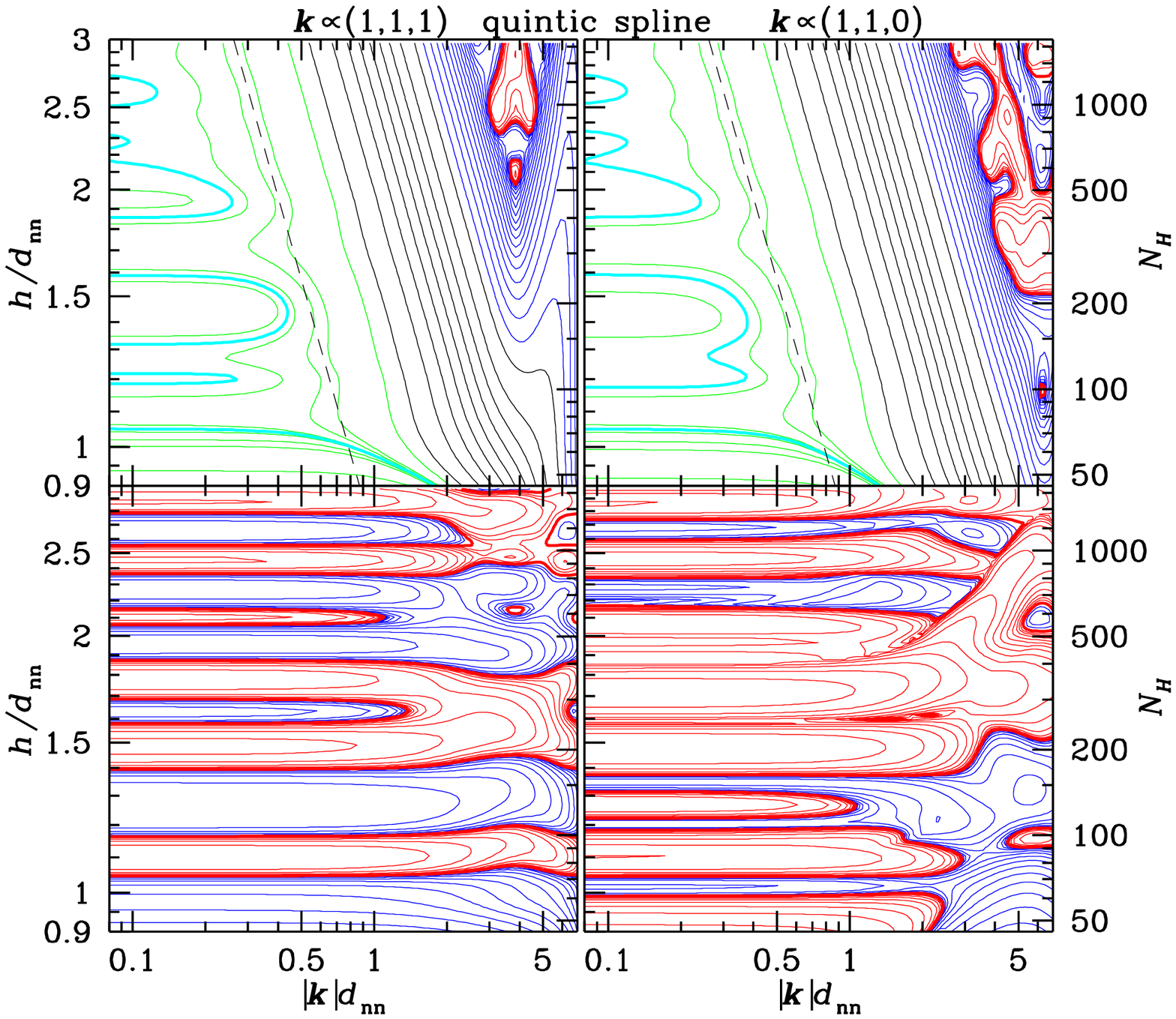}} \hfil
    \resizebox{86.25mm}{!}{\includegraphics{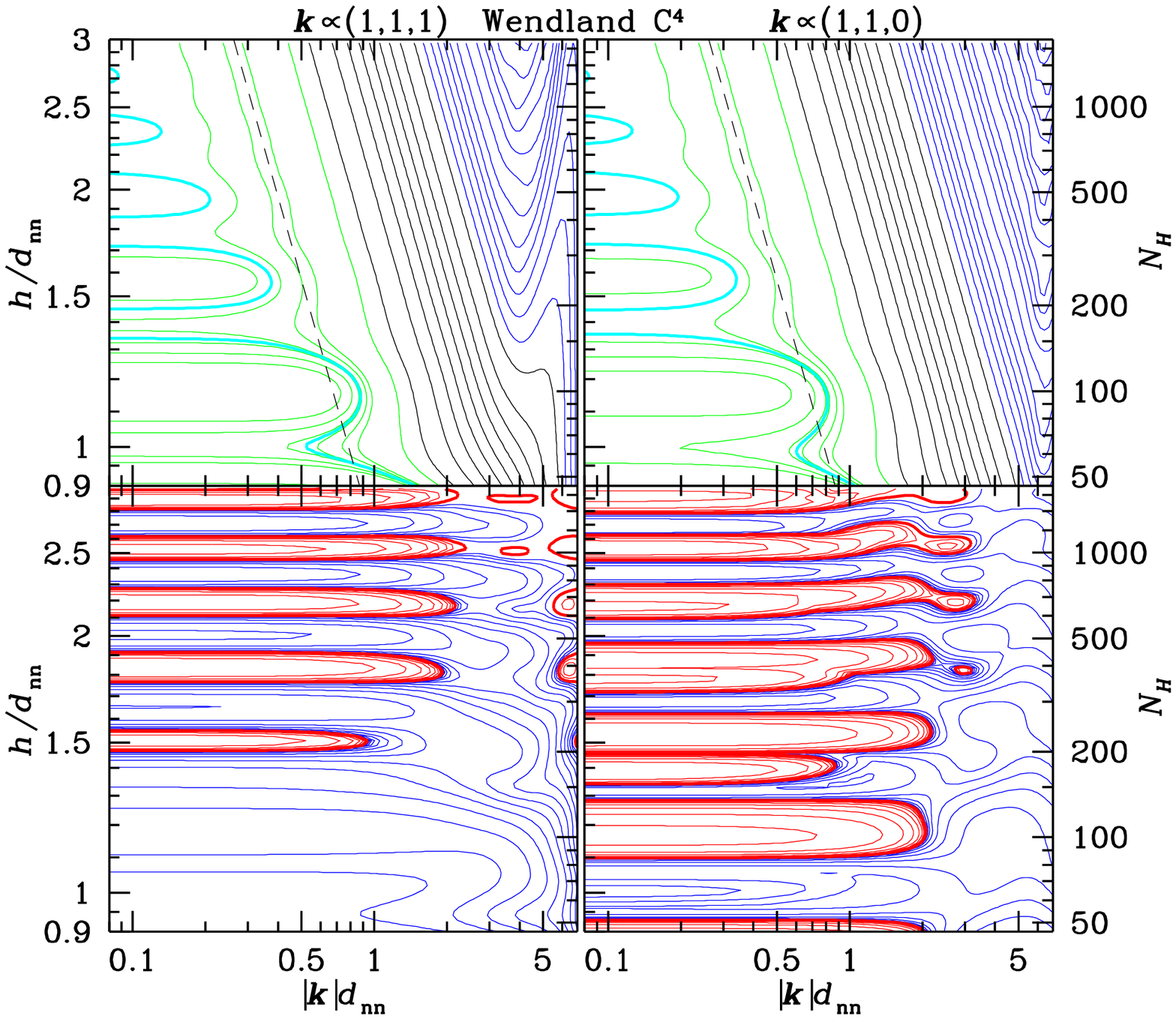}} \\[2ex]
    \resizebox{86.25mm}{!}{\includegraphics{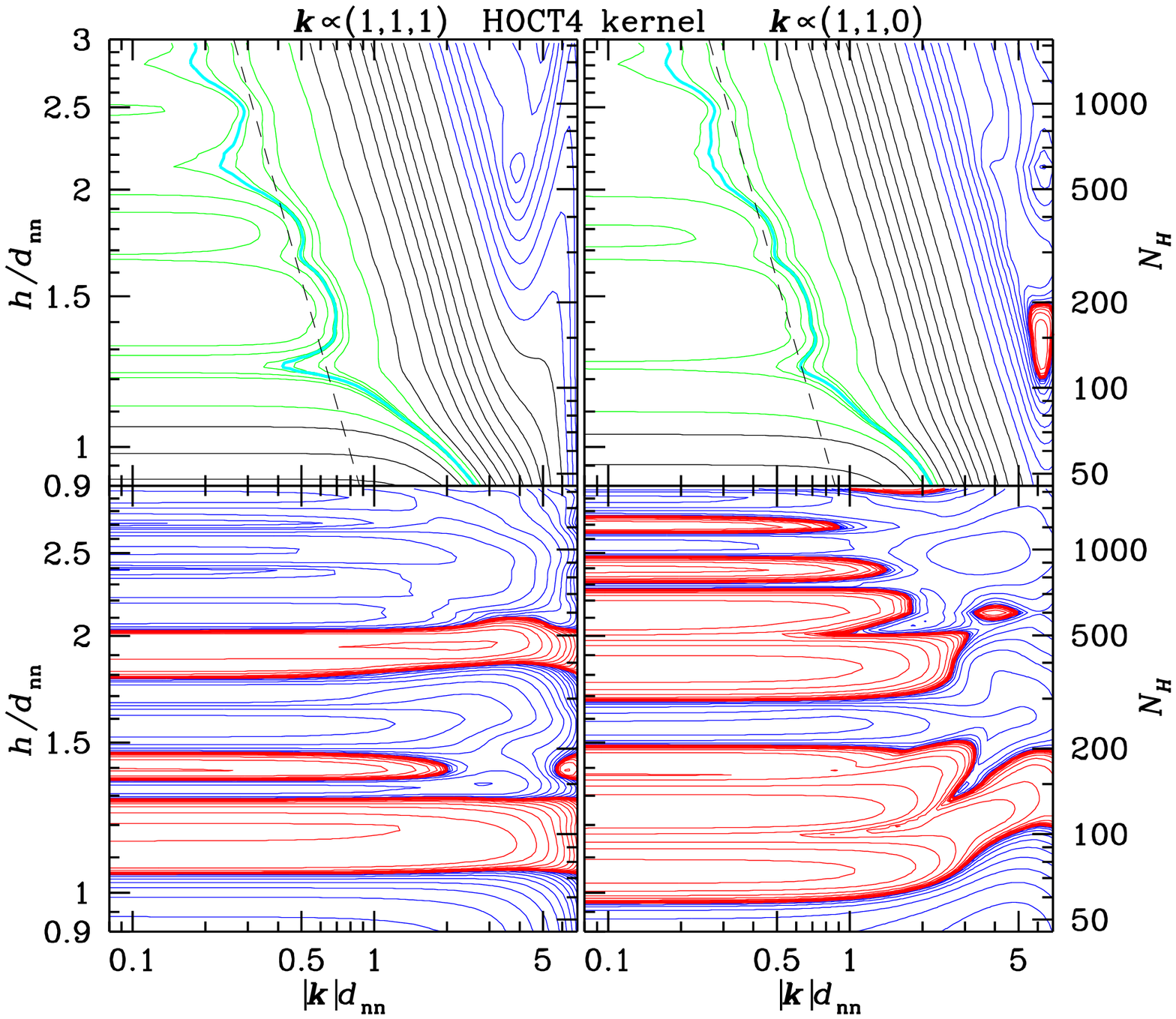}} \hfil
    \resizebox{86.25mm}{!}{\includegraphics{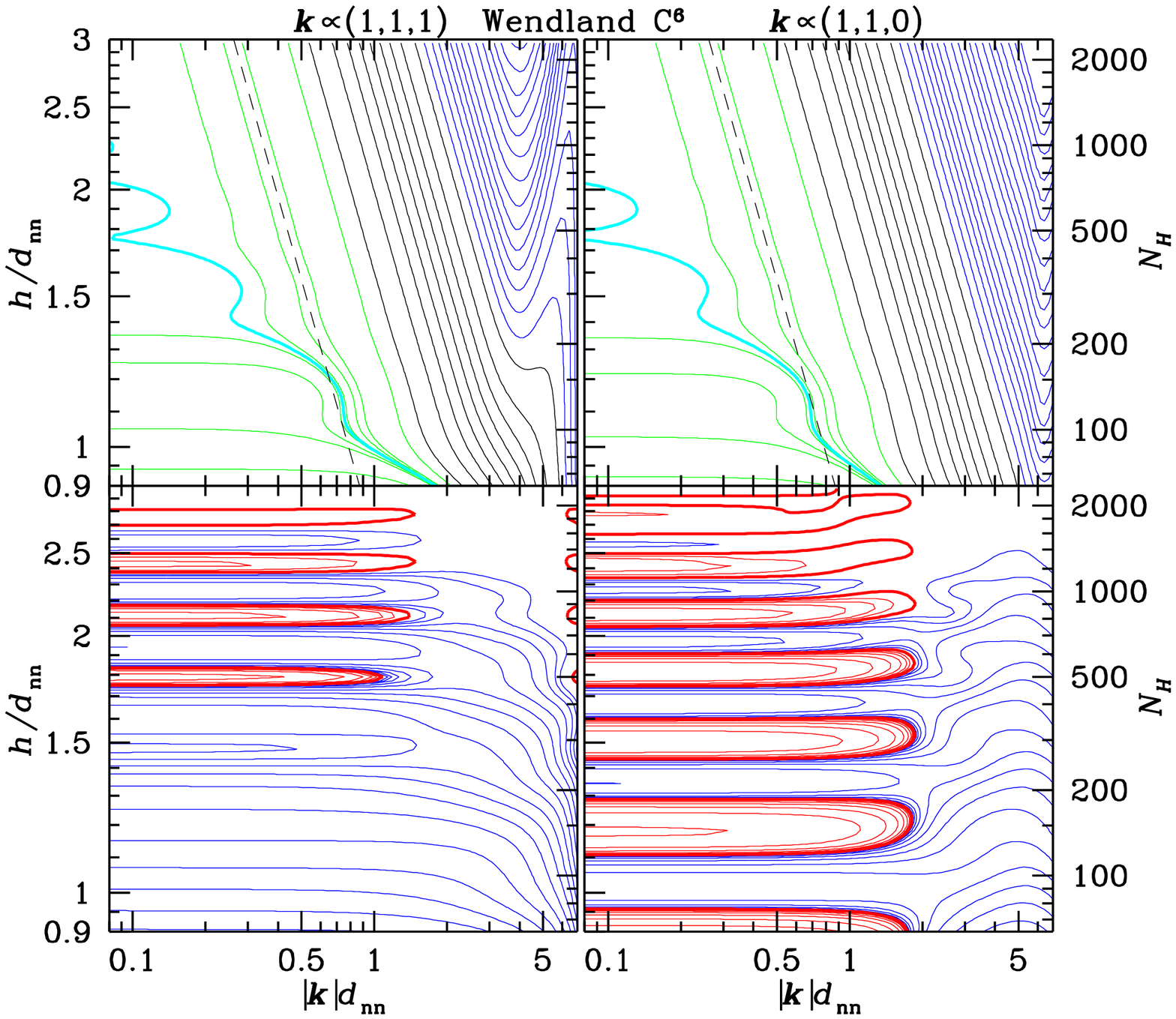}}
  \end{center}
  \vspace*{-3ex}
  \caption{
    \label{fig:stable:kernel}
    As Fig.~\ref{fig:stable:s3}, but for the quartic and quintic spline and the
    HOCT4 kernel (\emph{left}) and the Wendland $C^2$ to $C^6$ kernels
    (\emph{right}). The fine details of these contours are specific to the
    densest-sphere packing and will be different for more realistic glass-like
    particle distributions.}
\end{figure*}

The matrix $\B{\mathsf{P}}$ in equation (\ref{eq:P}) is not accessible to simple
interpretation. We will compute its eigenvalues for the various SPH kernels in
\S\!\S\ref{sec:stable:kern}-3 and Figs.~\ref{fig:stable:s3}-\ref{fig:disprel},
but first consider the limiting cases of the dispersion relation, allowing some
analytic insight.

\subsection{Limiting cases} \label{sec:stable:limit}
There are three spatial scales: the wavelength $\lambda=2\pi/|\B{k}|$, the
smoothing scale $h$, and the nearest neighbour distance $\dnn$. We will
separately consider the limit $\lambda\gg h$ of well resolved waves, the
continuum limit $h\gg\dnn$ of large neighbour numbers, and finally the combined
limit $\lambda\gg h\gg\dnn$.

\subsubsection{Resolved waves} \label{sec:stable:wave}
If $|\B{k}|h\ll1$, the argument of the trigonometric functions in equations
(\ref{eq:uUPi}a,b) is always small and we can Taylor expand them\footnote{In his
  analysis of 1D SPH, \cite{Rasio2000} also considers this simplification, but
  interprets it incorrectly as the limit $|\B{k}|\dnn\ll1$ regardless of
  $h$.}. If we also assume a locally isotropic particle distribution, this gives
to lowest order in $|\B{k}|$ ($\B{\mathsf{I}}$ is the unit matrix; see also
\S\ref{app:limit})
\begin{equation} \label{eq:P:resolved}
  \B{\mathsf{P}}\to\bar{c}^2\B{k}^{(2)}
  +\bar{\Xi}\,\frac{\bar{P}}{\bar{\rho}}
  \left[\left(\frac{2\nu}{\nu+2}-\{1\}\right)\B{k}^{(2)}
  +\frac{\nu\B{k}^2}{\nu+2}\,\B{\mathsf{I}}
  \right]
\end{equation}
with the eigenvalues
\begin{subequations}
  \begin{eqnarray}
    \frac{\ompar^2}{\B{k}^2} &=& \bar{c}^2 +
    \bar{\Xi}\,\frac{\bar{P}}{\bar{\rho}}
    \left(\frac{3\nu}{\nu+2}-\{1\}\right),
    \\
    \frac{\omega_{\perp1,2}^2}{\B{k}^2} &=& 
    \bar{\Xi}\,\frac{\bar{P}}{\bar{\rho}}
    \frac{\nu}{\nu+2}.
  \end{eqnarray}
\end{subequations}
The error of these relations is mostly dictated by the quality of the density
estimate, either directly via $\bar{\rho}$, $\bar{c}$, and $\bar{P}$, or
indirectly via $\bar{\Xi}$. The density correction method of equation
(\ref{eq:rho:corr}) can only help with the former, but not the latter. The
difference between constant and adapted $h$ is a factor 4/9 (for 3D) in favour
of the latter.

\subsubsection{Continuum limit} \label{sec:stable:cont}
For large neighbour numbers $\NH$, $H\gg\dnn$, $\bar{\Omega}\&\bar{\Pi}\to1$,
$\bar{\rho}\to\rho$ and the sums in equations\ (\ref{eq:uUPi}a,b) can be
approximated by integrals\footnote{Assuming a uniform particle distribution. A
  better approximation, which does not require $\dnn\ll H$, is to assume some
  \emph{radial distribution function} $g(r)$ (as in statistical mechanics of
  glasses) for the probability of any two particles having distance $r$. Such a
  treatment may well be useful in the context of SPH, but it is beyond the scope
  of our study.}
\begin{equation}
  \B{u} \to -\B{k}\,\widehat{W}(\B{k}),
  \;\;\text{and}\;\;
  \B{\mathsf{U}} \to \B{k}^{(2)}\,\widehat{W}(\B{k}),
\end{equation}
with $\widehat{W}(\B{k})$ the Fourier transform of $W(\B{x},h)$. Since
$\widehat{W}(\B{k})=\widehat{w}(H|\B{k}|)$, we have
$\partial\widehat{W}/\partial\ln h= \B{k}{\cdot}\B{\nabla}_{\!k}\widehat{W}$ and
thus from equation (\ref{eq:P})
\begin{equation} \label{eq:P:Fourier}
  \B{\mathsf{P}} \to c^2\B{k}^{(2)}\,
  \widehat{W}\left[\widehat{W}+\frac{2P}{\rho
      c^2}\bigg(1-\widehat{W}-
    \left\{\nu^{-1}\B{k}{\cdot}\B{\nabla}_{\!k}\widehat{W}\right\}
    \bigg)\right].
\end{equation}
$\widehat{W}(0)=1$, but towards larger $|\B{k}|$ the Fourier transform decays,
$\widehat{W}(\B{k})<1$, and in the limit $h|\B{k}|\gg1$ or $\lambda\ll h$,
$\widehat{W}\to0$: short sound waves are not resolved.

Negative eigenvalues of $\B{\mathsf{P}}$ in equation (\ref{eq:P:Fourier}), and
hence \corr{linear} instability, occur only if $\widehat{W}$ itself or the
expression within square brackets are negative. Since $\widehat{W}\le1$, the
latter can only happen if $P<0$, which does usually not arise in fluid
simulations (unless, possibly, one subtracts an average pressure), but possibly
in elasticity simulations of solids \citep*{GrayMonaghanSwift2001}, when it
causes the \emph{tensile} instability (an equivalent effect is present in
smoothed-particle MHD, see \citealt{PhillipsMonaghan1985,
  Price2012}). \cite{Monaghan2000} proposed an artificial repulsive short-range
force, effectuating an additional pressure, to suppress the tensile instability.

The pairing instability, on the other hand, is caused by $\widehat{W}<0$ for
some $H|\B{k}|>\kappa_0$. This instability can be avoided by choosing the
neighbour number $\NH$ small enough for the critical wave number $\kappa_0$ to
remain unsampled, i.e.\ $\kappa_0>\kappa_{\mathrm{Nyquist}}$ or
$H\lesssim\dnn\kappa_0/\pi$ (though such small $H$ is no longer consistent with
the continuum limit).

However, if the Fourier transform of the kernel is non-negative everywhere, the
pairing instability cannot occur for large $\NH$. As pairing is typically a
problem for large $\NH$, this suggests that kernels with $\widehat{W}(\B{k})>0$
for every $\B{k}$ are stable against pairing for \emph{all} values of $\NH$,
which is indeed supported by our results in \S\ref{sec:test:noise}.

\subsubsection{Resolved waves in the continuum limit} \label{sec:stable:long}
The combined limit of $\lambda\gg h\gg\dnn$ is obtained by inserting the Taylor
expansion (\ref{eq:w:Taylor}) of $\widehat{W}$ into equation
(\ref{eq:P:Fourier}), giving
\begin{equation} \label{eq:omega:large}
  \ompar^2 = c^2\B{k}^2\left(
  1+\sigma^2\B{k}^2
  \gamma^{-1}
  \big[\left\{ 2/\nu \right\}+1-\gamma\big]
  + \mathcal{O}(h^4|\B{k}|^4)
  \right).
\end{equation}
\cite{Monaghan2005} gave an equivalent relation for $\nu=1$ when the expression
in square brackets becomes $3-\gamma$ or $1-\gamma$ (for adapted or constant
$h$, respectively), which, he argues, bracket all physically reasonable values.
However, in 3D the value for adaptive SPH becomes $5/3-\gamma$,
i.e.\ \emph{vanishes} for the most commonly used adiabatic index.

In general, however, the relative error in the frequency is $\propto
\sigma^2\B{k}^2\propto(\sigma/\lambda)^2$. This shows that $h=2\sigma$ is indeed
directly proportional to the resolution scale, at least concerning sound waves.

\subsection{Linear stability of SPH kernels} \label{sec:stable:kern}
We have evaluated the eigenvalues $\ompar^2$ and $\omega_{\perp1,2}^2$ of the
matrix $\B{\mathsf{P}}$ in equation (\ref{eq:P}) for all kernels of
Table~\ref{tab:kernel}, as well as the HOCT4 and Gaussian kernels, for
unperturbed positions from densest-sphere packing (face-centred cubic
grid)\footnote{Avoiding an obviously unstable configuration, such as a cubic
  lattice, which may result in $\omega^2<0$ simply because the configuration
  itself was unstable, not the numerical scheme.}. In
Figs.~\ref{fig:stable:s3}\&\ref{fig:stable:kernel}, we plot the resulting
contours of $\omega^2\!/\!c^2\B{k}^2$ over wave number $|\B{k}|$ and smoothing
scale $h$ (both normalised by the nearest-neighbour distance $\dnn$) or $\NH$
\correct{ }on the right axes (except for the Gaussian kernel when $\NH$ is
ill-defined and we give $N_{\!h}$ instead) for two wave directions, one being
a nearest-neighbour direction.

\subsubsection{\corr{Linear stability against pairing}}
\label{sec:stable:pairing}
The top sub-panels of Figs.~\ref{fig:stable:s3}\&\ref{fig:stable:kernel} refer
to the longitudinal eigenvalue $\ompar^2$, when green and red contours are for,
respectively, $\ompar^2\approx c^2\B{k}^2$ and $\ompar^2<0$, the latter
indicative of the pairing instability. For the Gaussian kernel (truncated at
$16\sigma$; Fig.~\ref{fig:stable:s3}) $\ompar^2>0$ everywhere, proving its
stability\footnote{There is in fact $\ompar^2<0$ at values for $h$ larger than
  plotted. In agreement with our analysis in \S\ref{sec:stable:cont}, this is
  caused by truncating the Gaussian, which (like any other modification to avoid
  infinite neighbour numbers) invalidates the non-negativity of its Fourier
  transform. These theoretical results are confirmed by numerical findings of
  D.~Price (referee report), who reports pairing at large $h/\dnn$ for the
  truncated Gaussian.}, similar to the HOCT4 and, in particular the
higher-degree, Wendland kernels. In contrast, all the B-spline kernels obtain
$\ompar^2<0$ at sufficiently large $\NH$.

The quintic spline, Wendland $C^2$, and HOCT4 kernel each have a region of
$\ompar^2<0$ for $|\B{k}|$ close to the Nyquist frequency and $\NH\approx100$,
$\NH\approx40$, and $\NH\sim150$, respectively. In numerical experiments
similar to those described in \S\ref{sec:test:noise}, the corresponding
instability for the quintic spline and Wendland $C^2$ kernels can be triggered
by very small random perturbations to the grid equilibrium. However, such
modes are absent in glass-like configurations, which naturally emerge by
`cooling' initially random distributions. This strongly suggests, that these
kernel-$\NH$ combinations can be safely \correct{}used in practice. Whether
this also applies to the HOCT4 kernel at $\NH\sim150$ we cannot say, as we
have not run test simulations for this kernel. Note, that these islands of
linear instability at small $\NH$ are not in contradiction to the relation
between kernel Fourier transform and stability and are quite different from
the situation for the B-splines, which are only stable for sufficiently small
$\NH$.

\begin{figure}
  \begin{center}
    \resizebox{83mm}{!}{\includegraphics{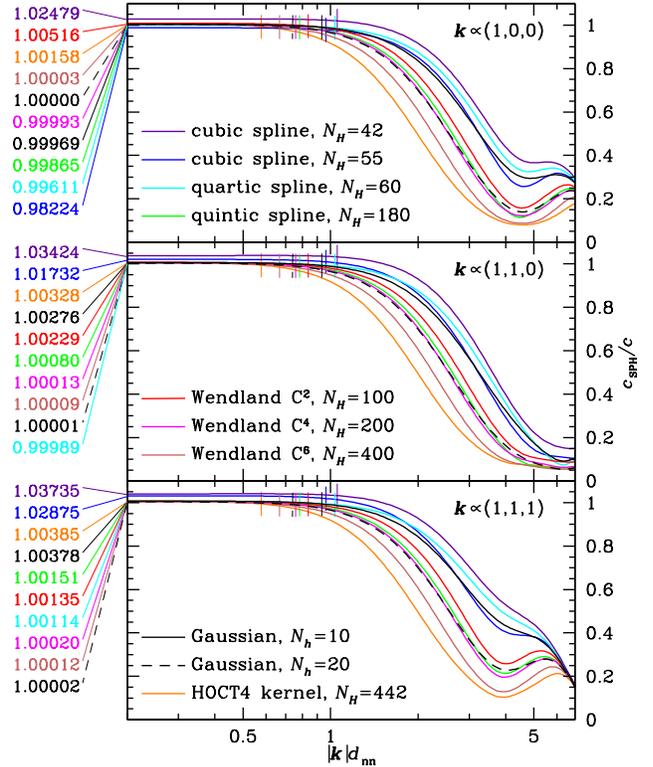}}
  \end{center}
  \vspace*{-3ex}
  \caption{
    \label{fig:disprel}
    Ratio of SPH sound speed $c_{\mathrm{SPH}}=\ompar/|\B{k}|$ for the equation
    of state $P\propto\rho^{5/3}$ to the correct $c$ for the kernel-$\NH$
    combinations of Table~\ref{tab:NH:kern} and three different wave directions
    (as indicated) for particles in densest-sphere packing with
    nearest-neighbour separation $\dnn$. These curves are horizontal cuts
    through the top panels of
    Figs.~\ref{fig:stable:s3}\&\ref{fig:stable:kernel}. The thin vertical lines
    indicate sound with wavelength $\lambda=8h$, corresponding to the dashed
    lines in Figs.~\ref{fig:stable:s3}\&\ref{fig:stable:kernel}.  }
\end{figure}
\subsubsection{\corr{Linear transverse instability?}}
\label{sec:stable:transverse}
The bottom sub-panels of Figs.~\ref{fig:stable:s3}\&\ref{fig:stable:kernel}
show $\omega_{\perp2}^2/c^2\B{k}^2$, when both families of kernels have
$0\approx|\omega_{\perp1,2}^2|\ll c^2\B{k}^2$ with either sign
occurring. $\omega_{\perp1,2}^2<0$ implies growing transverse
modes\footnote{\cite{ReadHayfieldAgertz2010} associate $\omega_{\perp1,2}^2<0$
  with a `banding instability' which appeared near a contact discontinuity in
  some of their simulations. However, they fail to provide convincing
  arguments for this connection, as their stability analysis is compromised by
  the use of the unstable cubic lattice.}, which we indeed found in
simulations starting from a slightly perturbed densest-sphere
packing. However, such modes are not present in glass-like configurations,
\correct{which strongly} suggests, that transverse modes are not a problem in
practice.

\subsection{Numerical resolution of sound waves}
\label{sec:stable:sound}
The dashed lines in Figs.~\ref{fig:stable:s3}\&\ref{fig:stable:kernel} indicate
sound with wavelength $\lambda=8h$. For $h\gtrsim\dnn$, such sound waves are
well resolved in the sense that the sound speed is accurate to
$\lesssim1\%$. This is similar to grid methods, which typically require about
eight cells to resolve a wavelength.

The effective SPH sound speed can be defined as
$c_{\mathrm{SPH}}=\ompar/|\B{k}|$. In Fig.~\ref{fig:disprel} we plot the ratio
between $c_{\mathrm{SPH}}$ and the correct sound speed as function of wave
number for three different wave directions and the ten kernel-$\NH$ combinations
of Table~\ref{tab:NH:kern} (which also gives their formal resolutions). The
transition from $c_{\mathrm{SPH}}\approx c$ for long waves to
$c_{\mathrm{SPH}}\ll c$ for short waves occurs at $|\B{k}|\dnn\gtrsim1$, but
towards longer waves for larger $h/\dnn$, as expected.

For resolved waves ($\lambda\gtrsim8h$: left of the thin vertical lines in
Fig.~\ref{fig:disprel}), $c_{\mathrm{SPH}}$ obtains a value close to $c$, but
with clear differences between the various kernel-$\NH$
combinations. Surprisingly, the standard cubic spline kernel, which is used
almost exclusively in astrophysics, performs very poorly with errors of few
percent, for both $\NH=42$ and 55. This is in stark contrast to the quartic
spline with similar $\NH=60$ but $c_{\mathrm{SPH}}$ accurate to
$<1\%$. Moreover, the quartic spline with $\NH=60$ resolves shorter waves better
than the cubic spline with a smaller $\NH=55$, in agreement with
Table~\ref{tab:NH:kern}.

We should note that these results for the numerical sound speed assume a
perfectly smooth simulated flow. In practice, particle disorder \corr{degrades}
the performance, in particular for smaller $\NH$\corr{, and the resolution of
  SPH is limited by the need to suppress this degradation via increasing $h$
  (and $\NH$).}

\begin{table}
  \begin{center}
    \begin{tabular}{@{}lrrrl@{}}
      \multicolumn{1}{c}{kernel} & 
      $\NH$ & $N_{\!h}$ & $h/\dnn$ & $h(\hat{\rho}/m)^{1/3}$ \\[-0.5ex]
      \hline \\[-3ex]
      cubic spline   &  42      &  6.90 & 1.052 & 1.181\\
      cubic spline   &  55      &  9.04 & 1.151 & 1.292\\
      quartic spline &  60      &  7.29 & 1.072 & 1.203\\
      quintic spline & 180      & 17.00 & 1.421 & 1.595\\
      Wendland $C^2$ & 100      & 13.77 & 1.325 & 1.487\\
      Wendland $C^4$ & 200      & 18.58 & 1.464 & 1.643\\
      Wendland $C^6$ & 400      & 27.22 & 1.662 & 1.866\\
      HOCT4          & 442      & 42.10 & 1.923 & 2.158\\
      Gaussian       & $\infty$ & 10.00 & 1.191 & 1.337\\
      Gaussian       & $\infty$ & 20.00 & 1.500 & 1.684\\
    \end{tabular}
  \end{center}
  \vspace*{-1ex}
  \caption{Some quantities (defined in \S\ref{sec:smooth:scale}) for
    kernel-$\NH$ combinations used in Fig.~\ref{fig:disprel} and the test
    simulations of \S\ref{sec:test}. $\dnn$ is the nearest-neighbour distance
    for densest-sphere packing, which has number density
    $n=\!\sqrt{2}\,\dnn^{-3}$. The cubic spline with $\NH\approx42$ is the most
    common choice in astrophysical simulations, the other $\NH$ values for the
    B-spline are near the pairing-stability boundary, hence obtaining close to
    the greatest possible reduction \corr{of the `E$_0$ errors'}. For $\NH=100$,
    200, 400, we picked the Wendland kernel which gave best results for the
    vortex test of \S\ref{sec:test:vortex}.
    \label{tab:NH:kern}}
\end{table}
\section{Test simulations} \label{sec:test}
In order to assess the Wendland kernels and compare them to the standard
B-spline kernels in practice, we present some test simulations which emphasise
the pairing, \corr{strong shear}, and shocks. All these simulations are
done in 3D using periodic boundary conditions, $P= K\rho^{5/3}$, conservative
SPH (equation~\ref{eq:hydro}), and the \cite{CullenDehnen2010} artificial
viscosity treatment, which invokes dissipation only for compressive flows, and
an artificial conductivity similar to that of \cite{ReadHayfield2012}.  For some
tests we used various values of $\NH$ per kernel, but mostly those listed in
Table~\ref{tab:NH:kern}.

\subsection{Pairing in practice} \label{sec:test:noise}
In order to test our theoretical predictions regarding the pairing instability,
we evolve noisy initial conditions with 32000 particles until equilibrium is
reached. Initially, $\dot{\B{x}}_i=0$, while the initial $\B{x}_i$ are generated
from densest-sphere packing by adding normally distributed offsets with (1D)
standard deviation of one unperturbed nearest-neighbour distance $\dnn$. To
enable a uniform-density equilibrium (a glass), we suppress viscous heating.

\begin{figure}
  \begin{center}
    \resizebox{38mm}{!}{\includegraphics{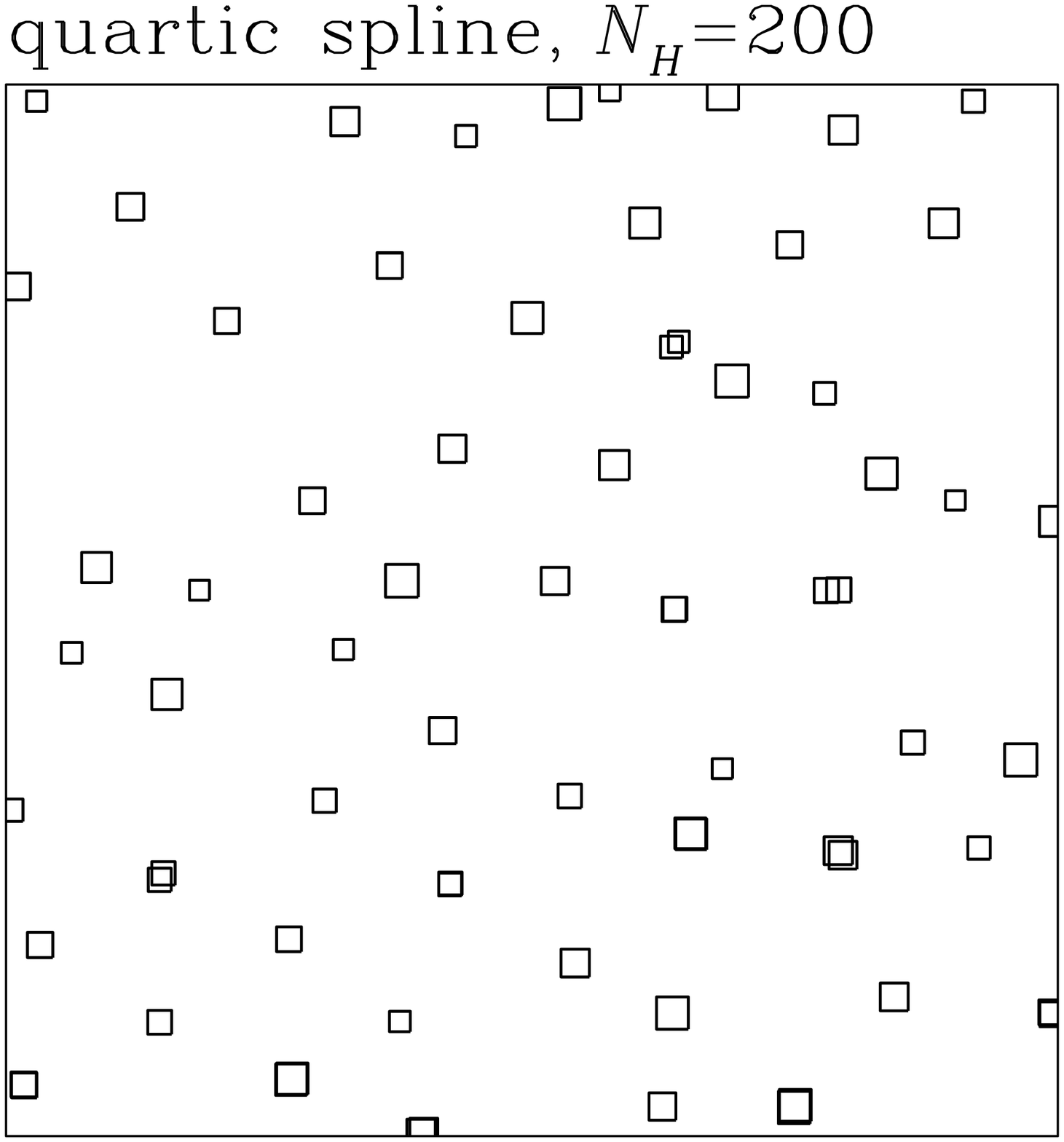}}\hfil
    \resizebox{38mm}{!}{\includegraphics{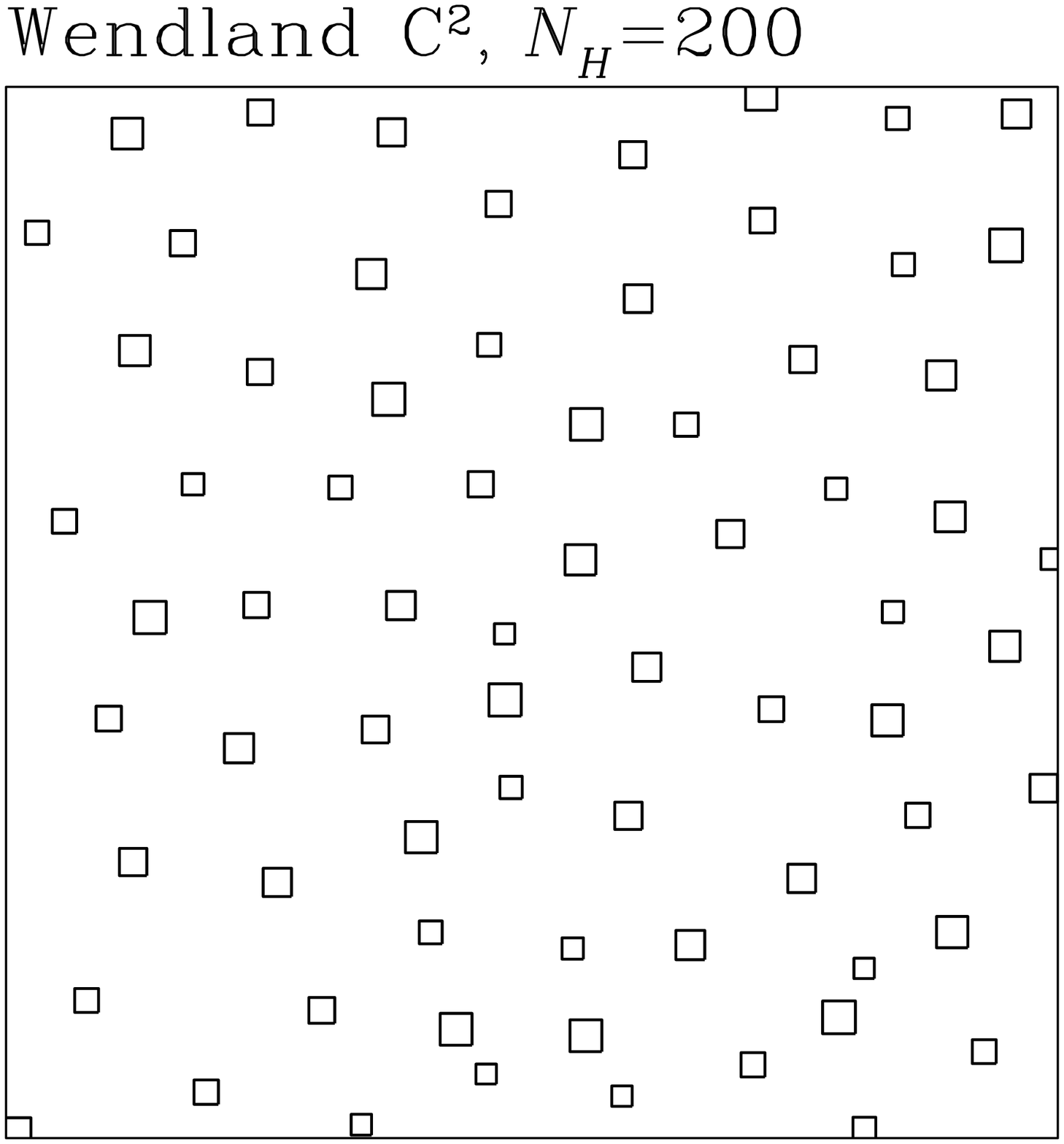}}
  \end{center}
  \vspace*{-3ex}
  \caption{
    \label{fig:noise:xy}
    Final $x$ and $y$ positions for particles at $|z|<\dnn/2$ for the tests of
    \S\ref{sec:test:noise}. Symbol size is linear in $z$: two overlapping
    symbols of the same size indicate particle pairing.}
\end{figure}

The typical outcome of these simulations is either a glass-like configuration
(right panel of Fig.~\ref{fig:noise:xy}) or a distribution with particle pairs
(left panel of Fig.~\ref{fig:noise:xy}). In order to quantify these outcomes, we
compute for each particle the ratio
\begin{equation} \label{eq:rmin}
  r_{\min,i} = \min_{j\neq i}\big\{|\B{x}_i-\B{x}_{\!j}|\big\}/H_i
\end{equation}
between its actual nearest-neighbour distance and kernel-support radius. The
maximum possible value for $r_{\min}$ occurs for densest-sphere packing, when
$|\B{x}_i-\B{x}_{\!j}|\ge\dnn=(n/\!\sqrt{2})^{1/3}$ with $n$ the number
density. Replacing $\hat{\rho}_i$ in equation (\ref{eq:NH}) with $m_in$, we
obtain
\begin{equation} \label{eq:rmin:max}
  r_{\min} \lesssim \big(3\NH/2^{5/2}\pi\big)^{1/3}.
\end{equation}
Thus, the ratio 
\begin{equation} \label{eq:qmin}
  q_{\min,i} = \frac{r_{\min,i}}{(3\NH/2^{5/2}\pi)^{1/3}} \approx
  \min_{j\neq i} \left\{
  \frac{|\B{x}_i-\B{x}_{\!j}|}{d_{\mathrm{nn,grid}}}\right\}
\end{equation}
is an indicator for the regularity of the particle distribution around particle
$i$. It obtains a value very close to one for perfect densest-sphere packing and
near zero for pairing, while a glass typically gives $q_{\min,i}\sim0.7$.

\begin{figure}
  \begin{center}
    \resizebox{78mm}{!}{\includegraphics{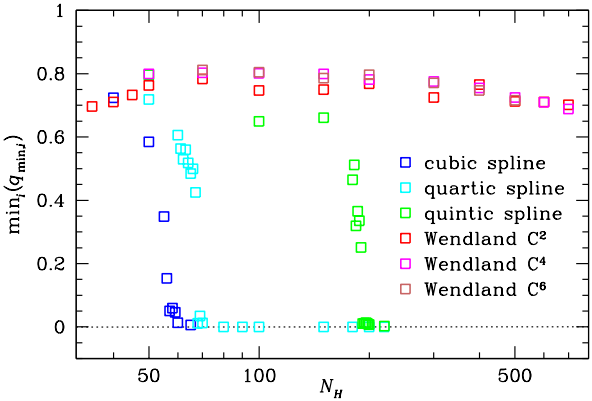}}
  \end{center}
  \vspace*{-3ex}
  \caption{
    \label{fig:noise:qmin}
    Final value of the overall minimum of $q_{\min,i}$ (equation~\ref{eq:qmin})
    for simulations starting from noisy initial conditions for all kernels of
    Table~\ref{tab:kernel} (same colour coding as in Fig.~\ref{fig:rho}; using
    the density correction of \S\ref{sec:kern:dens} for the Wendland kernels) as
    function of $\NH$. For densest-sphere packing $q_{\min,i}\approx1$, for a
    glass $0\ll q_{\min,i}<1$, while $q_{\min,i}\sim0$ indicates particle
    pairing.  }
\end{figure}

Fig.~\ref{fig:noise:qmin} plots the final value for the overall minimum of
$q_{\min,i}$ for each of a set of simulations. For all values tested for $\NH$
(up to 700), the Wendland kernels show no indication of a single particle
pair. This is in stark contrast to the B-spline kernels, all of which suffer
from particle pairing. The pairing occurs at $\NH>67$ and 190 for the quartic,
and quintic spline, respectively, whereas for the cubic spline
$\min_i\{q_{\min,i}\}$ approaches zero more gradually, with
$\min_i\{q_{\min,i}\}\le0.16$ at $\NH > 55$. These thresholds match quite well
the suggestions of the linear stability analysis in
Figs.~\ref{fig:stable:s3}\&\ref{fig:stable:kernel} (except that the indications
of instability of the quintic spline at $\NH\approx100$ and the Wendland $C^2$
kernel at $\NH\approx40$ are not reflected in our tests here). The quintic (and
higher-order) splines are the only option amongst the B-spline kernels for $\NH$
appreciably larger than ${\sim\,}50$.

We also note that $\min_i\{q_{\min,i}\}$ \corr{grows} substantially faster, in
particularly early on, for the Wendland kernels than for the B-splines,
especially when operating close to the stability boundary.

\begin{figure*}
  \begin{center}
    \resizebox{46.70mm}{!}{\includegraphics{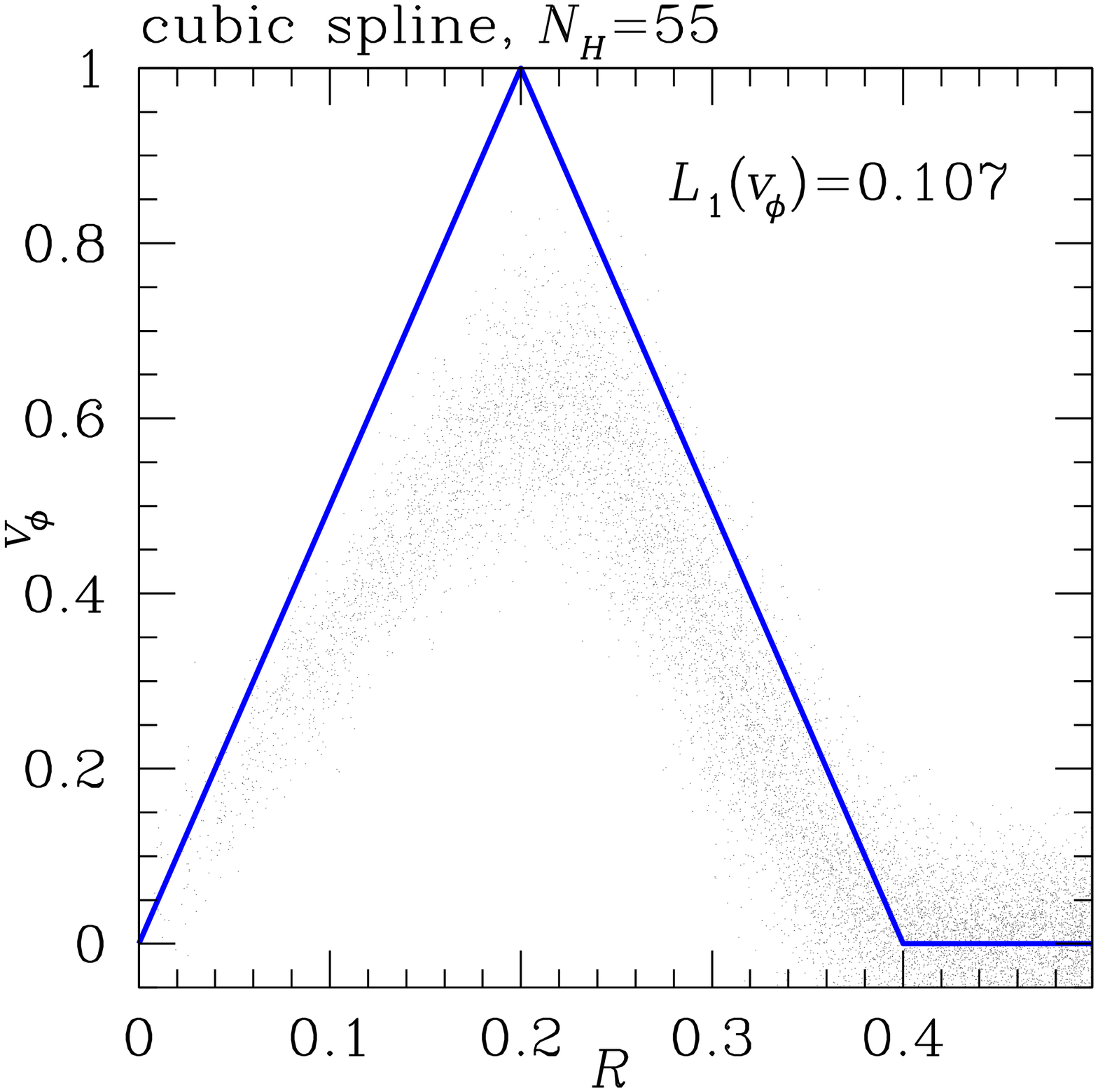}}
    \resizebox{41.55mm}{!}{\includegraphics{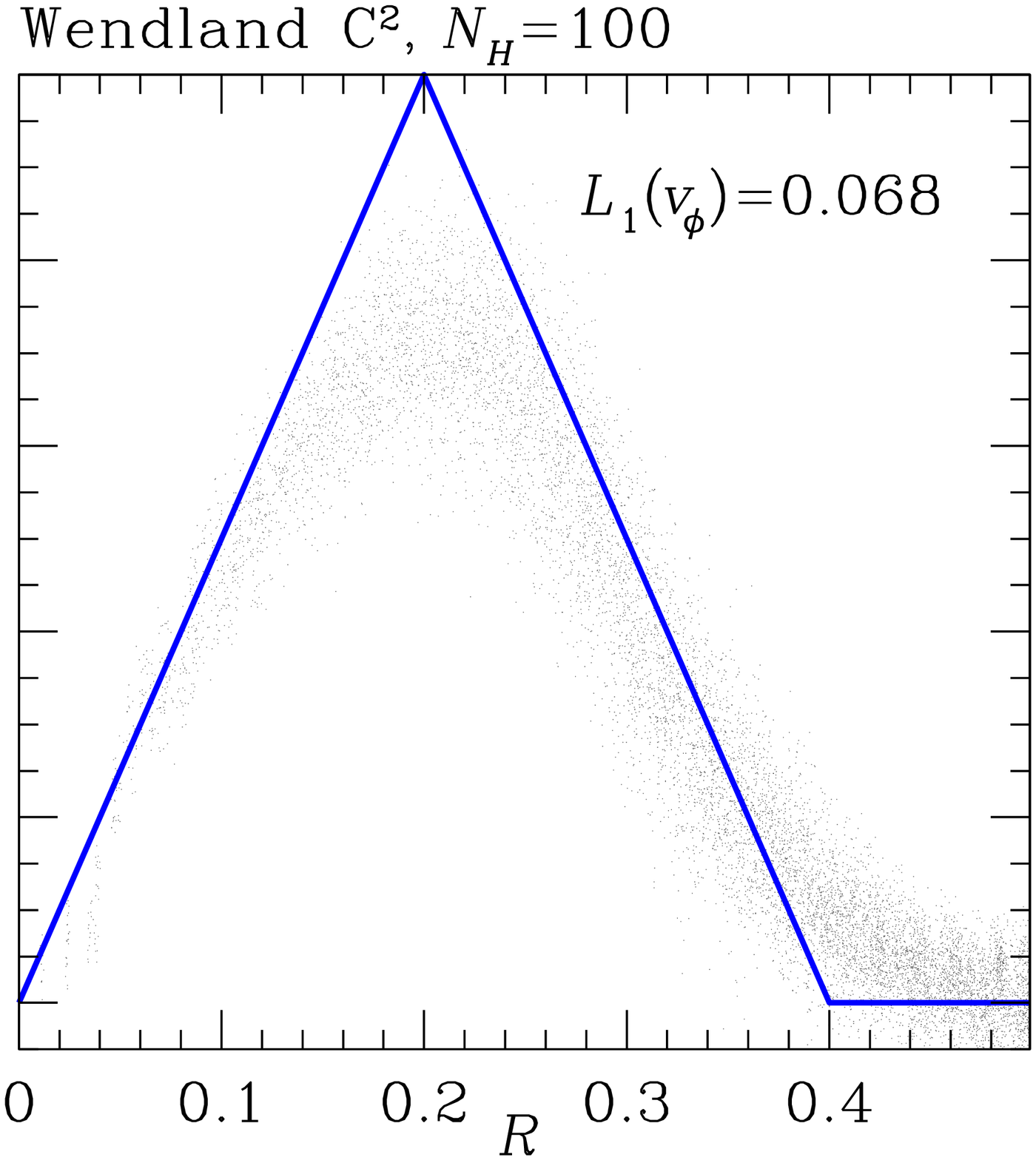}}
    \resizebox{41.55mm}{!}{\includegraphics{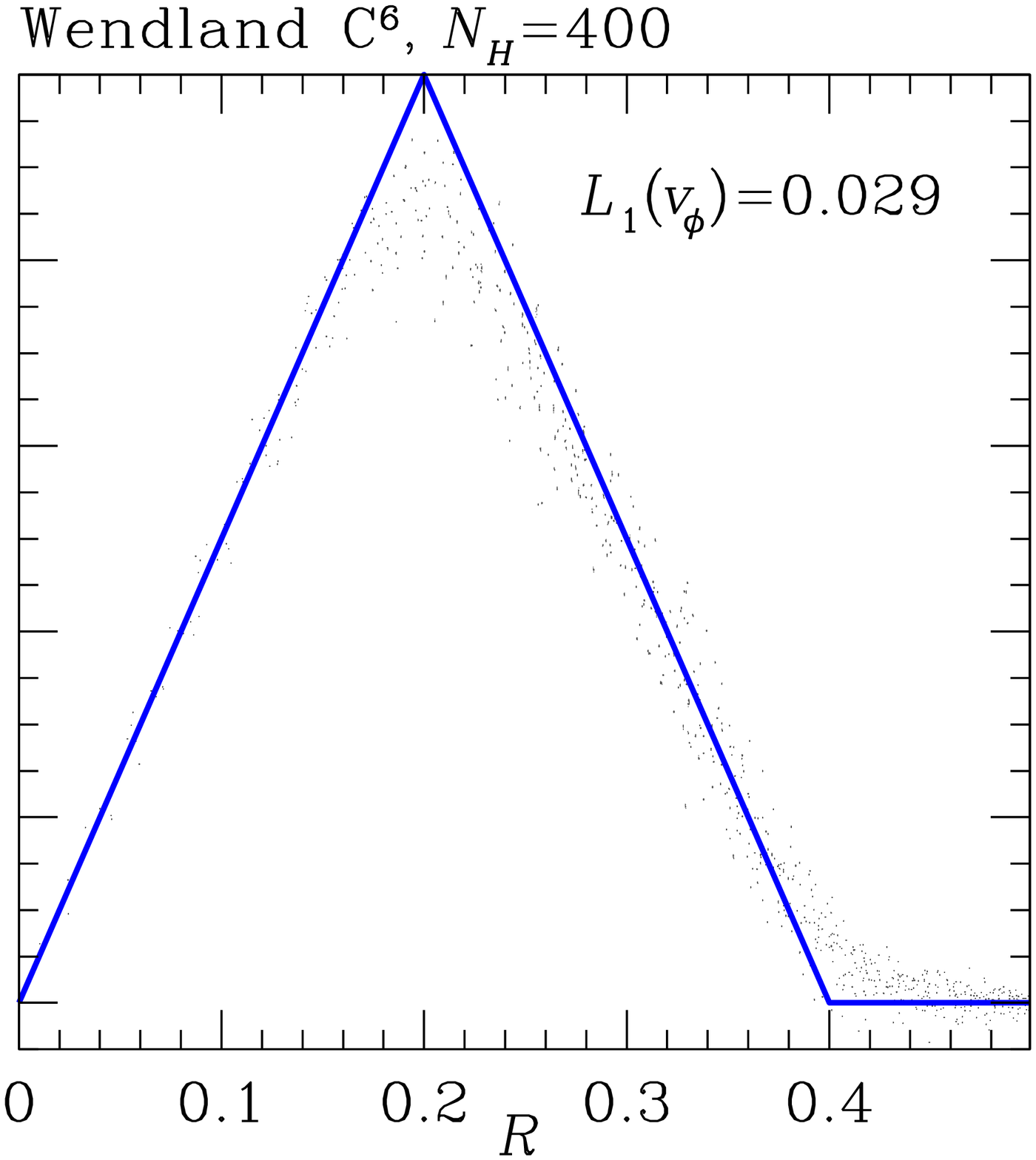}}
    \resizebox{41.55mm}{!}{\includegraphics{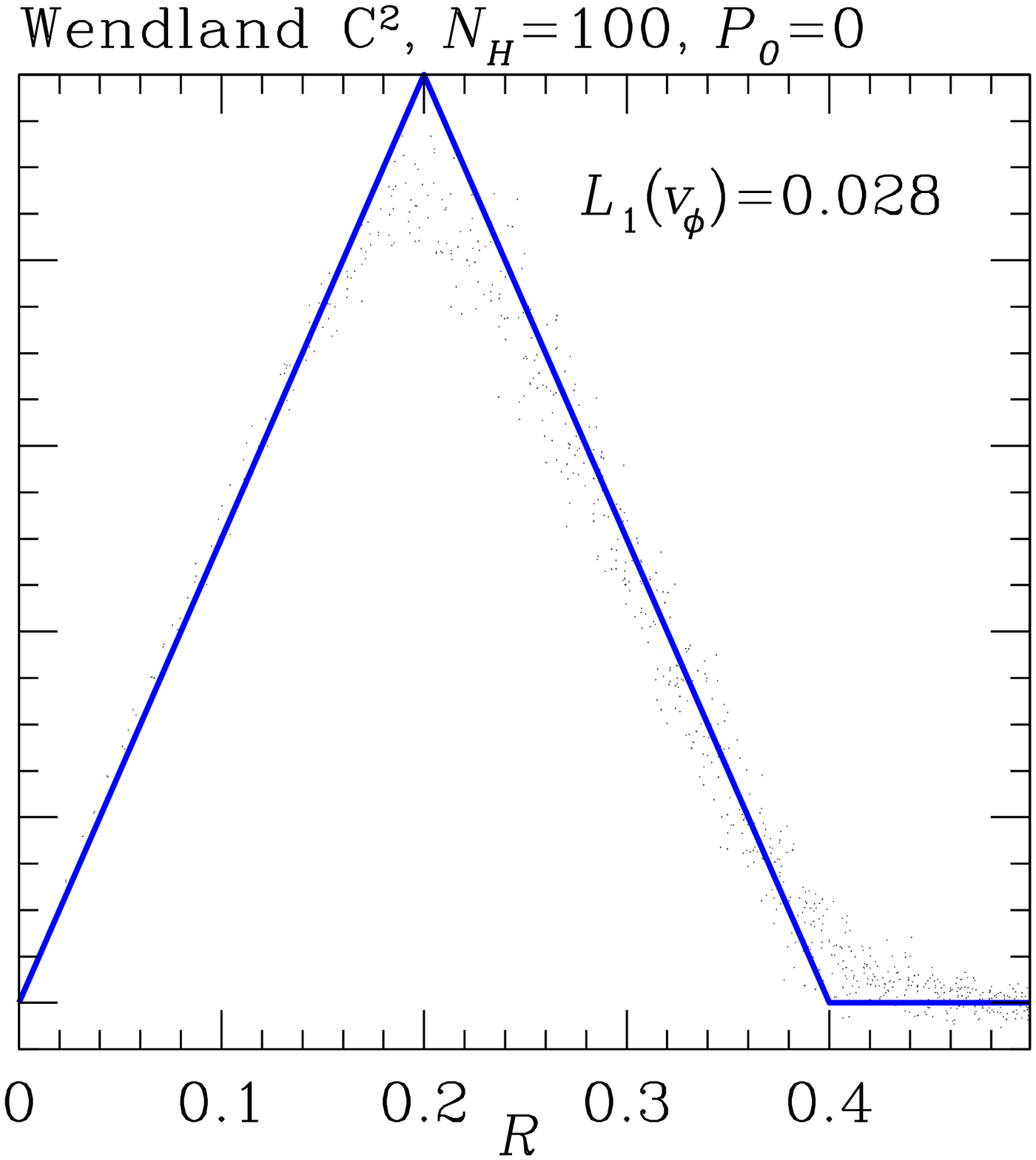}}
  \end{center}
  \vspace*{-3ex}
  \caption{
    \label{fig:gresho:dv}
    Velocity profiles 
    the Gresho-Chan vortex test at time $t=1$ with the lowest resolution of
    $N_{\mathrm{1D}}=50.8$ for three different kernel-$\NH$ pairs (as indicated)
    and one simulation (\emph{right panel}) with $P_0=0$ in equation
    (\ref{eq:gresho:P}), when the `\corr{E$_0$} error' is naturally much
    weaker. Only particles at $|z|<0.05$ are plotted.}
\end{figure*}
\subsection{The Gresho-Chan vortex test} \label{sec:test:vortex}
As discussed in the introduction, particle disorder is unavoidably generated in
shearing flows, inducing `\corr{E$_0$} errors' in the forces and causing
modelling errors. A critical test of this situation consists of a differentially
rotating fluid of uniform density in centrifugal balance
(\citealt{GreshoChan1990}, see also \citealt{LiskaWendroff2003},
\citealt{Springel2010:review}, and \citealt{ReadHayfield2012}). The pressure and
azimuthal velocity are
\begin{subequations}
  \label{eqs:gresho}
  \begin{eqnarray}
    \label{eq:gresho:P}
    P     &=& \left\{\begin{array}{l@{\quad\text{for}\;\;}lllll}
    P_0+12.5 R^2                      &0   &\le &R &< &0.2, \\
    P_0+12.5 R^2+4-20R+4\ln(5R) &0.2 &\le &R &< &0.4, \\
    P_0+2 (2\ln2-1)                 &0.4 &\le &R &; &    \end{array}\right.
    \\
    \label{eq:gresho:v}
    v_\phi &=& \left\{\begin{array}{l@{\quad\text{for}\;\;}lllll}
    5R     &0   &\le &R &< &0.2, \\
    2-5R &0.2 &\le &R &< &0.4, \\
    0      &0.4 &\le &R &  &    \end{array}\right.
  \end{eqnarray}
\end{subequations}
with $P_0=5$ and $R$ the cylindrical radius. We start our simulations from
densest-sphere packing with effective one-dimensional particle numbers
$N_{\mathrm{1D}}=51$, 102, 203, or 406. The initial velocities and pressure are
set as in equations (\ref{eqs:gresho}).

There are three different causes for errors in this test. First, an overly
viscous method reduces the differential rotation, as shown by
\cite{Springel2010:review}; this effect is absent from our simulations owing to
the usage of the \cite{CullenDehnen2010} dissipation switch. Second, the
`\corr{E$_0$} error' generates noise in the velocities which in turn triggers
some viscosity. Finally, finite resolution implies that the sharp velocity kinks
at $R=0.2$ and 0.4 cannot be fully resolved (in fact, the initial conditions are
not in SPH equilibrium because the pressure gradient at these points is smoothed
such that the SPH acceleration is not exactly balanced with the centrifugal
force).

In Fig.~\ref{fig:gresho:dv} we plot the azimuthal velocity at time $t=1$ for a
subset of all particles at our lowest resolution of $N_{\mathrm{1D}}=51$ for
four different kernel-$\NH$ combinations. The leftmost is the standard cubic
spline with $\NH=55$, which considerably suffers from \corr{particle disorder
  and hence} `\corr{E$_0$} errors (but also obtains too low $v_\phi$ at
$R<0.2$).

The second is the Wendland $C^2$ kernel with $\NH=100$, which still suffers from
the `\corr{E$_0$} error'. The last two are for the Wendland $C^6$ kernel with
$\NH=400$ and the Wendland $C^2$ kernel with $\NH=100$ but with $P_0=0$ in
equation (\ref{eq:gresho:P}). In both cases, the `\corr{E$_0$} error' is much
reduced (and the accuracy limited by resolution) either because of large
neighbour number or because of a reduced pressure.

\begin{figure}
  \begin{center}
    \resizebox{75mm}{!}{\includegraphics{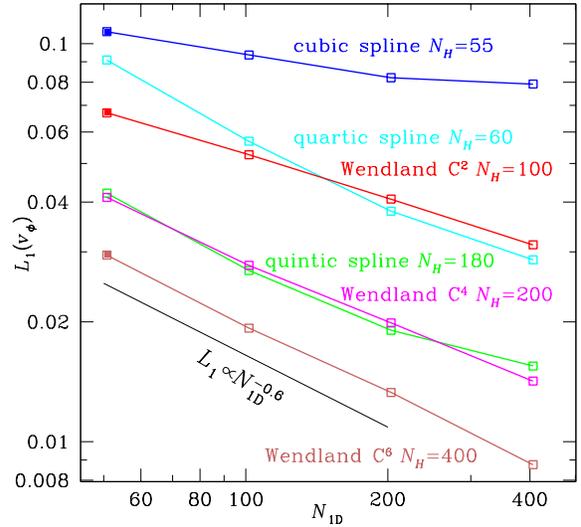}}
  \end{center}
  \vspace*{-3ex}
  \caption{
    \label{fig:gresho}
    Convergence of the $L_1$ velocity error for the Gresho-Chan test with
    increasing number of particles (for a cubic lattice $N_{\mathrm{1D}}$ equals
    the number of cells along one side of the computational domain) for various
    kernel-$\NH$ combinations (those with filled squares are shown in
    Fig.~\ref{fig:gresho:dv}). A comparison with Fig.~6 of Read \& Hayfield
    (2012) shows that the Wendland $C^6$ kernel with $\NH=400$ performs better
    than the HOCT4 kernel with $\NH=442$.  }
\end{figure}
\begin{figure*}
  \begin{center}
    \resizebox{36.30mm}{!}{\includegraphics{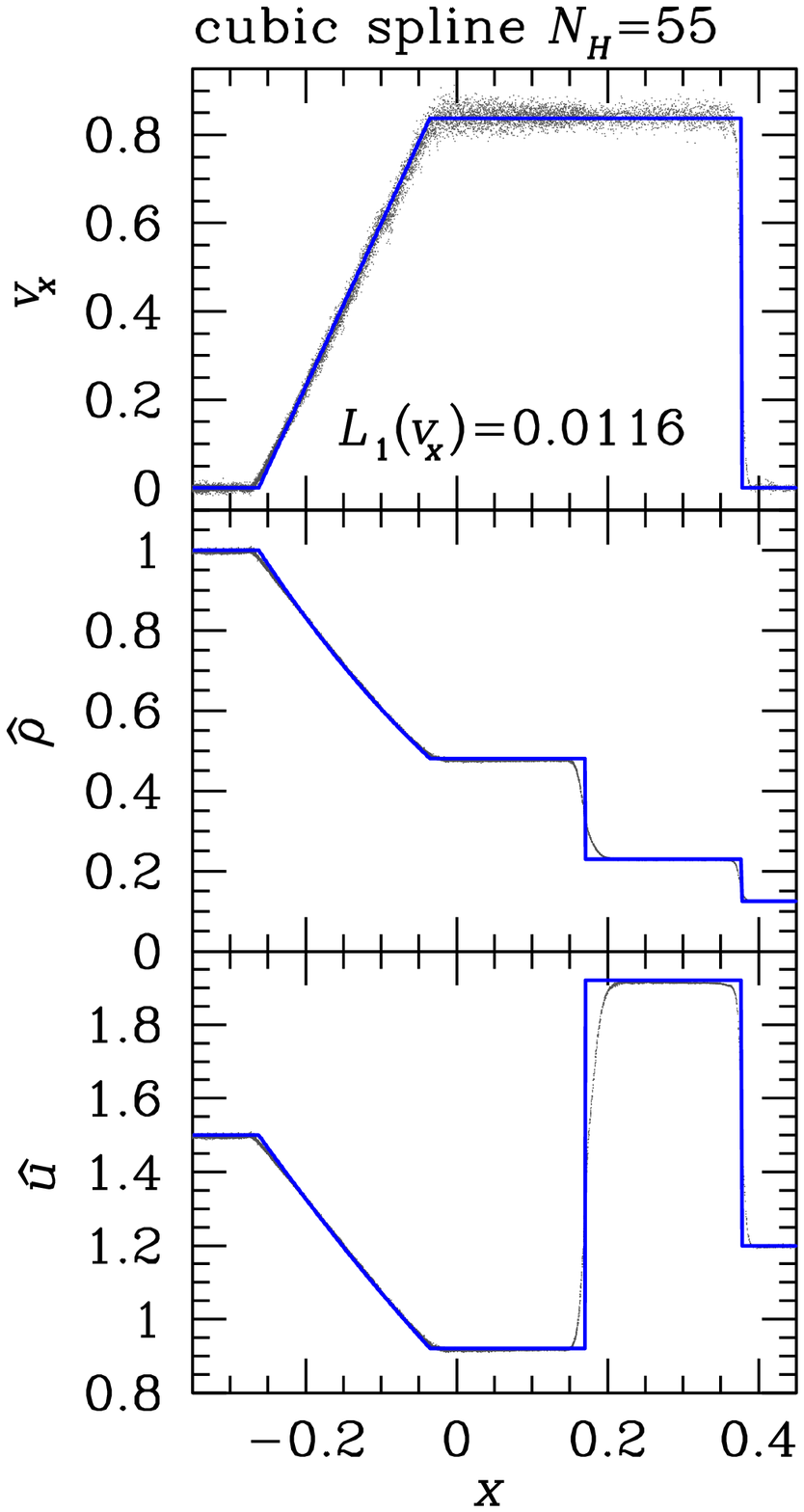}}\!
    \resizebox{27.90mm}{!}{\includegraphics{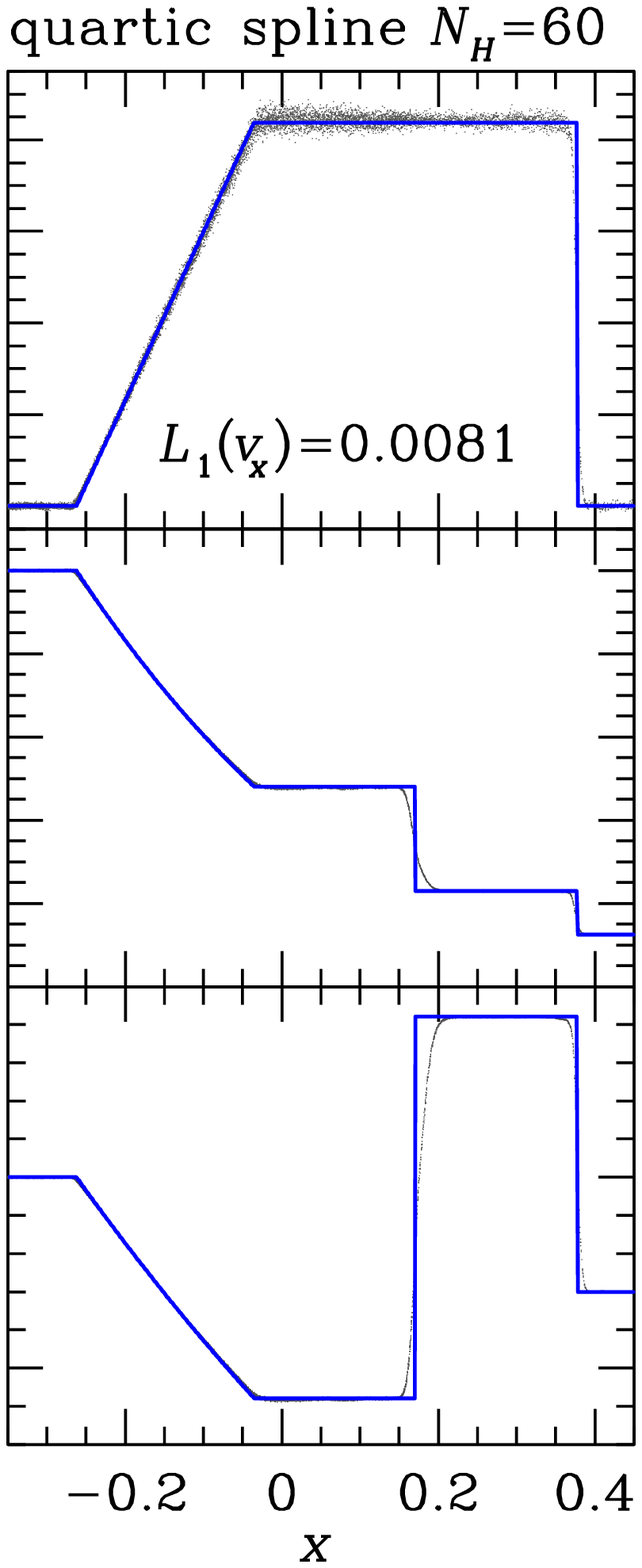}}\!
    \resizebox{27.90mm}{!}{\includegraphics{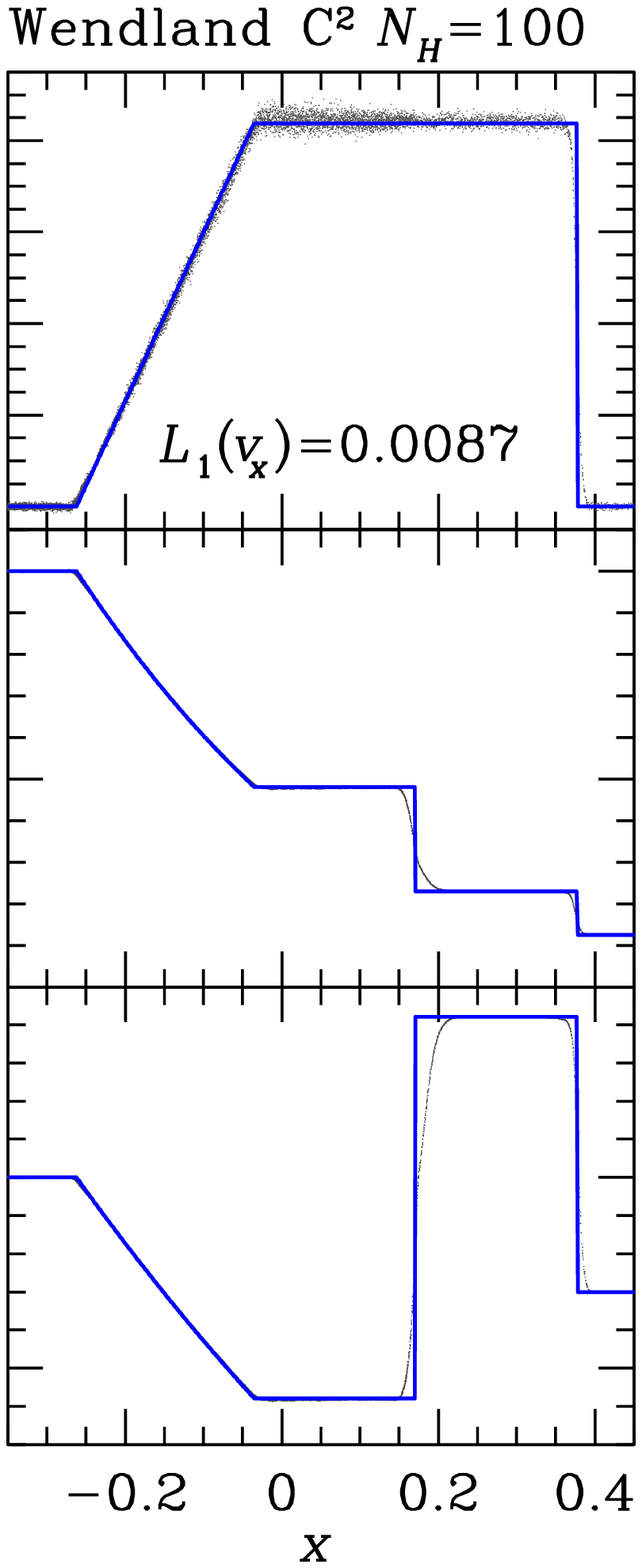}}\!
    \resizebox{27.90mm}{!}{\includegraphics{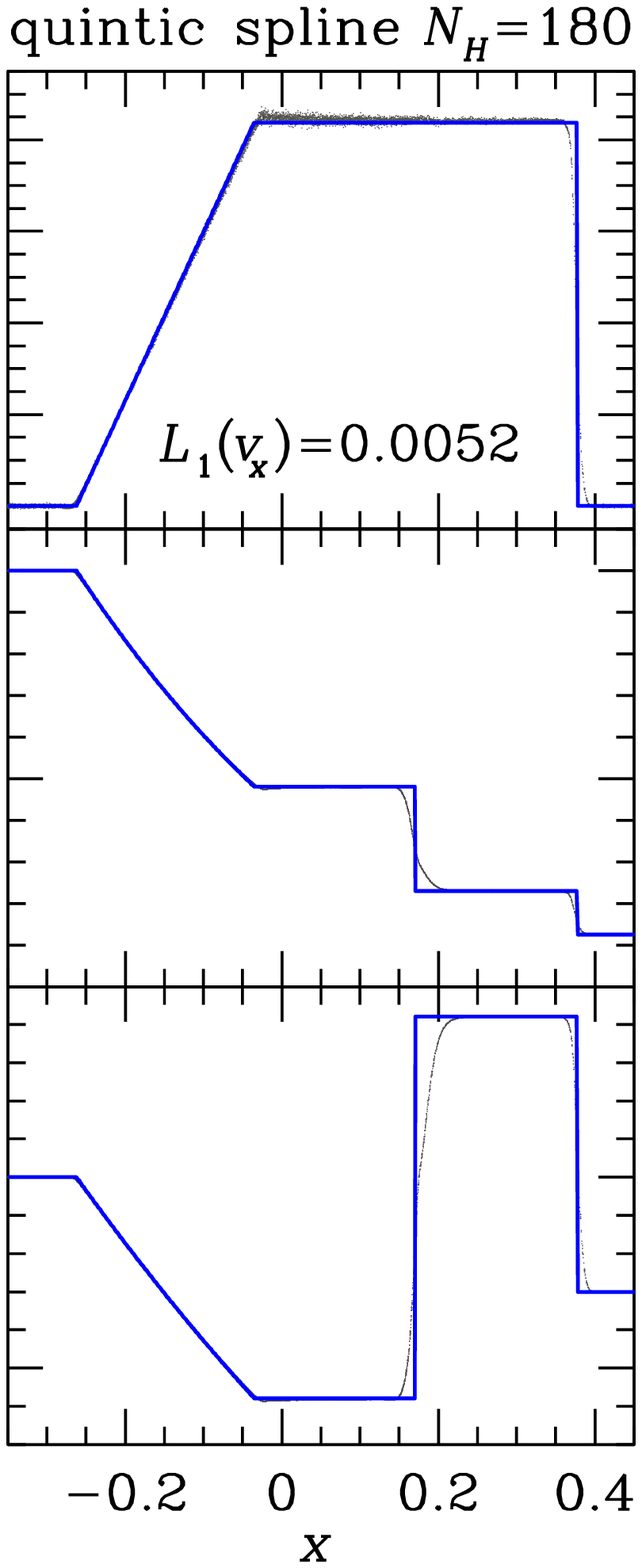}}\!
    \resizebox{27.90mm}{!}{\includegraphics{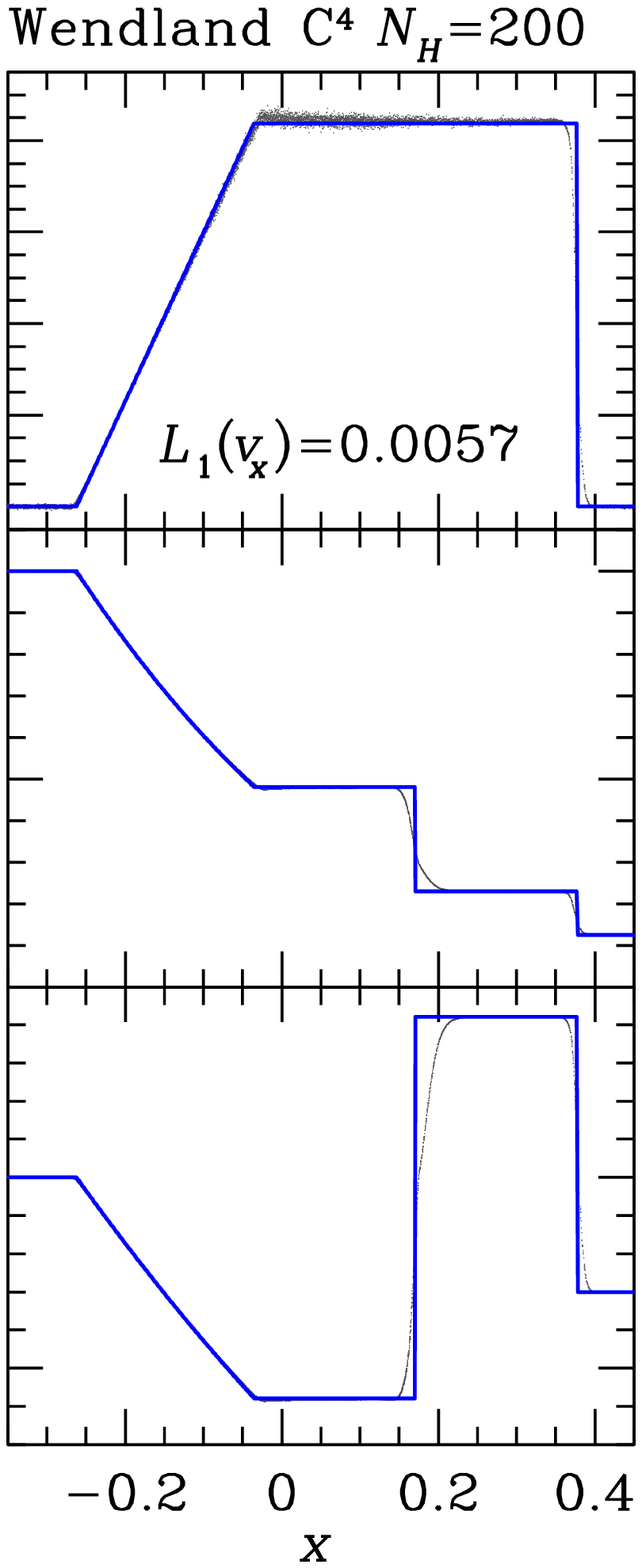}}\!
    \resizebox{27.90mm}{!}{\includegraphics{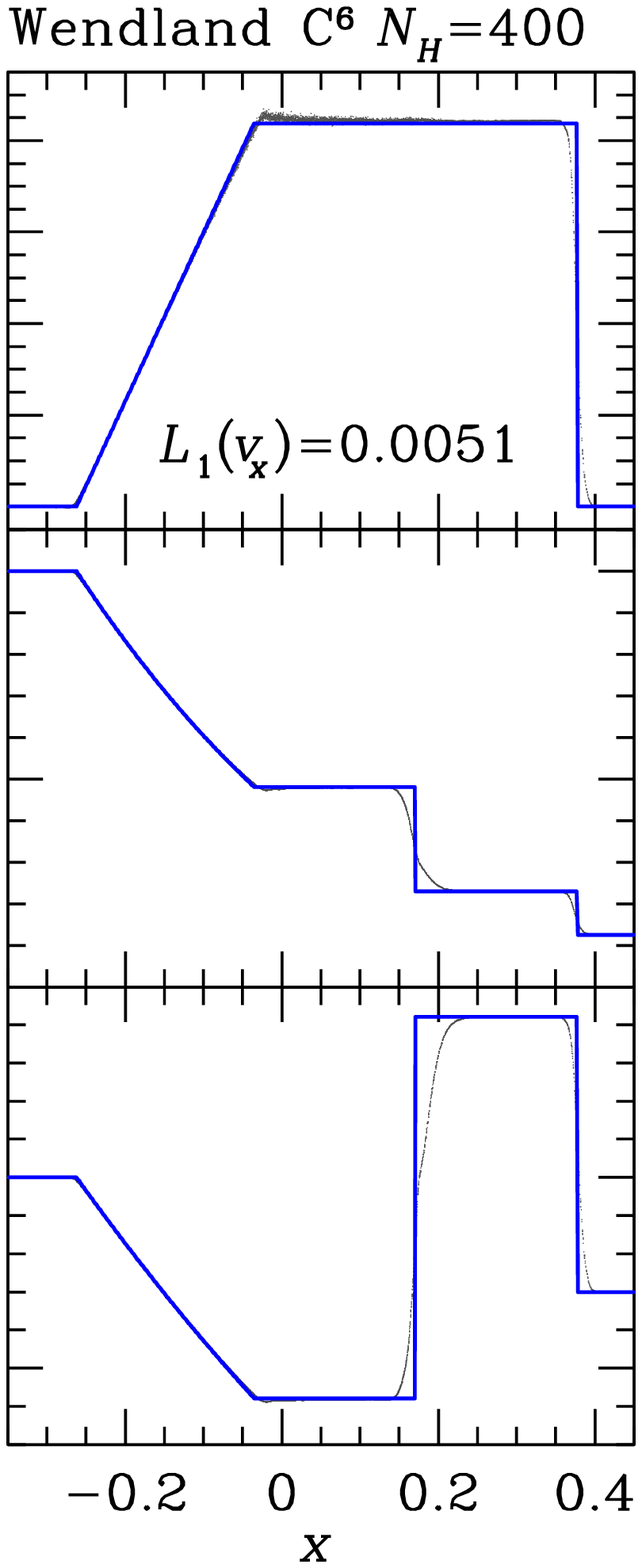}}
  \end{center}
  \vspace*{-3ex}
  \caption{
    \label{fig:sod}
    Velocity, density, and thermal energy profiles for the shock tube test at
    $t=0.2$ for the same kernel-$\NH$ combinations as in
    Fig.~\ref{fig:gresho} in order of increasing $\NH$ (\emph{points}:
    particles at $|y|,|z|<0.005$) and the exact solutions (\emph{solid}). The
    $L_1$ velocity errors reported are computed for the range $-0.4<x<0.5$.}
\end{figure*}

In Fig.~\ref{fig:gresho}, we plot the convergence of the $L_1$ velocity error
with increasing numerical resolution for all the kernels of
Table~\ref{tab:kernel}, but with another $\NH$ for each, see also
Table~\ref{tab:NH:kern}. For the B-splines, we pick a large $\NH$ which still
gives sufficient stability against pairing, while for $\NH=100$, 200, and 400 we
show the Wendland kernel that gave best results.  For the cubic spline, the
results agree with the no-viscosity case in Fig.~6 of
\cite{Springel2010:review}, demonstrating that our dissipation switch
effectively yields inviscid SPH. We also see that the rate of convergence (the
slope of the various curves) is lower for the cubic spline than any other
kernel. This is caused by systematically too low $v_\phi$ in the rigidly
rotating part at $R<0.2$ (see leftmost panel if Fig.~\ref{fig:gresho:dv}) at all
resolutions. The good performance of the quartic spline is quite surprising, in
particular given the rather low $\NH$. The quintic spline at $\NH=180$ and the
Wendland $C^4$ kernel at $\NH=200$ obtain very similar convergence, but are
clearly topped by the Wendland $C^6$ kernel at $\NH=400$, demonstrating that
high neighbour number is really helpful in strong shear flows.

\subsection{Shocks} \label{sec:test:shoc}
Our final test is the classical \cite{Sod1978} shock tube, a 1D Riemann problem,
corresponding to an initial discontinuity in density and pressure.  Unlike most
published applications of this test, we perform 3D simulations with glass-like
initial conditions. Our objective here is (1) to verify the \corr{`E$_0$-error'}
reductions at larger $\NH$ and (2) the resulting trade-off with the resolution
across the shock and contact discontinuities. Other than for the vortex tests of
\S\ref{sec:test:vortex}, we only consider one value for the number $N$ of
particles but the same six kernel-$\NH$ combinations as in
Fig.~\ref{fig:gresho}. The resulting profiles of velocity, density, and thermal
energy are plotted in Fig.~\ref{fig:sod} together with the exact solutions.

Note that the usual over-shooting of the thermal energy near the contact
discontinuity (at $x=0.17$) is prevented by our artificial conductivity
treatment. This is not optimised and likely over-smoothes the thermal energy
(and with it the density). However, here we concentrate on the velocity.

For the cubic spline with $\NH=55$, there is significant velocity noise in the
post-shock region. This is caused by the re-ordering of the particle positions
after the particle distribution becomes anisotropically compressed in the
shock. This type of noise is a well-known issue with multi-dimensional SPH
simulations of shocks \citep[e.g.][]{Springel2010:review,Price2012}. With
increasing $\NH$ the velocity noise is reduced, but because of the smoothing of
the velocity jump at the shock (at $x=0.378$) the $L_1$ velocity error does not
approach zero for large $\NH$.

Instead, for sufficiently large $\NH$ ($\gtrsim200$ in this test), the $L_1$
velocity error saturates: any \corr{`E$_0$-error'} reduction for larger
$\NH$ is balanced by a loss of resolution. The only disadvantage of larger $\NH$
is an increased computational cost (by a factor $\sim1.5$ when moving from the
quintic spline with $\NH=180$ to the Wendland kernel $C^6$ with $\NH=400$, see
Fig.~\ref{fig:time}).

\section{Discussion} \label{sec:disc}
\subsection{What causes the pairing instability?} \label{sec:disc:clum}
The Wendland kernels have an inflection point and yet show no signs of the
pairing instability. This clearly demonstrates that the traditional ideas for
the origin of this instability (\'a la \citealt{SwegleHicksAttaway1995}, see the
introduction) were incorrect. Instead, our linear stability analysis shows that
in the limit of large $\NH$ pairing is caused by a negative kernel Fourier
transform $\widehat{W}$, whereas the related tensile instability with the same
symptoms is caused by an (effective) negative pressure.  While it is intuitively
clear that negative pressure causes pairing, the effect of $\widehat{W}<0$ is
less obvious.  Therefore, we now provide another explanation, not restricted to
large $\NH$.

\subsubsection{
  \corr{The pairing instability as artifact of the density estimator}}
\label{sec:pairing:alt}
By their derivation from the Lagrangian (\ref{eq:L}), the SPH forces
$m_i\ddot{\B{x}}_i=\partial\mathcal{L}/\partial\B{x}_i
=-\partial\hat{U}\!/\partial\B{x}_i$ tend to reduce the estimated total thermal
energy $\hat{U}=\sum_im_i\,u(\hat{\rho}_i,s_i)$ at fixed entropy\footnote{This
  holds, of course, also for an isothermal gas, when $u$ is constant, but not
  the entropy $s$, so that $(\partial u/\partial \rho)_s\neq 0$.}. \corr{Thus,
  hydrostatic equilibrium corresponds to an extremum of $\hat{U}$, and stable
  equilibrium to a minimum when small positional changes meet opposing for\-ces.
  Minimal $\hat{U}$ is obtained for uniform $\hat{\rho}_i$}, since a
re-distribution of the particles in the same volume but with a spread of
$\hat{\rho}_i$ gives larger $\hat{U}$ (assuming uniform $s_i$). \corr{An
  equilibrium is meta-stable, if $\hat{U}$ is only a local (but not the global)
  minimum. Several extrema can occur if different particle distributions, each
  obtaining (near\mbox{-})\discretionary{}{}{}uniform $\hat{\rho}_i$, have
  different average $\hat{\rho}$.} Consider, for example, particles in
densest-sphere packing, replace each by a pair and increase the spacing by
$2^{1/3}$, so that the \corr{average} density \corr{$\rho$ (but not
  $\hat{\rho}$)} remains unchanged. This fully paired distribution is in
equilibrium with uniform $\hat{\rho}_i$, but the \corr{\emph{effective}
  neighbour number is reduced by a factor 2} (for the same smoothing scale). Now
if $\hat{\rho}(\NH/2)<\hat{\rho}(\NH)$, the paired distribution has lower
$\hat{U}$ than the original and is favoured.

In practice (and in our simulations in \S\ref{sec:test:noise}), the pairing
instability appears gradually: for $\NH$ just beyond the stability boundary,
only few particle pairs form and \corr{the effective reduction of $\NH$ is by a
  factor $f\le2$}. We conclude, therefore, that
\begin{center}
  \noindent\textbf{\boldmath pairing occurs if
    $\hat{\rho}(\corr{\NH/f})<\hat{\rho}(\NH)$ for \corr{some
    $1<f\le2$}.}
\end{center}
From Fig.~\ref{fig:rho} we see that for the B-spline kernels $\hat{\rho}(\NH)$
always has a minimum and hence \corr{satisfies our condition}, while this never
occurs for the Wendland or HOCT4 kernels\footnote{The density correction of
  \S\ref{sec:kern:dens} does not affect these arguments, because during a
  simulation $\epsilon$ in equation (\ref{eq:rho:corr}) is \emph{fixed} and in
  terms of our considerations here the solid curves in Fig.~\ref{fig:rho} are
  simply lowered by a constant.}. The \corr{stability} boundary (between squares
and crosses \corr{in Fig.~\ref{fig:rho}}) is towards slightly larger $\NH$ than
\corr{the minimum of $\hat{\rho}(\NH)$, indicating $f>1$ (but also note that the
  curves} are based on a regular grid \corr{instead of a glass as the squares}).

\subsubsection{
  \corr{The relation to `\corr{E$_0$} errors' and particle re-ordering}}
A disordered particle distribution is typically not in equilibrium, but has
non-uniform $\hat{\rho}_i$ and hence non-minimal $\hat{U}$. The SPH forces, in
particular the\corr{ir} `\corr{E$_0$} errors' \corr{(which occur even for
  constant pressure)}, then drive the evolution towards smaller $\hat{U}$ and
hence equilibrium with either a glass-like order or pairing \corr{\citep[see
    also][\S5]{Price2012}}. Thus, the minimisation of $\hat{U}\!$ is the
underlying driver for both the particle \corr{re-}ordering capability of SPH and
the pairing instability. This also means that when operating near the stability
boundary, for example using $\NH=55$ for the cubic spline, this re-ordering is
much reduced. This is why in Fig.~\ref{fig:noise:qmin} the transition between
glass and pairing is not abrupt: for $\NH$ just below the stability boundary the
glass-formation, which relies on the re-ordering mechanism, is very slow and not
finished by the end of our test simulations.

A\corr{n immediate} corollary of these considerations is that \corr{any SPH-like
  method without `\corr{E$_0$} errors' does not have an automatic re-ordering
  mechanism. This applies to modifications of the force equation that avoid the
  `\corr{E$_0$} error', but also to the} method of \cite{HessSpringel2010},
which \corr{employs} a Voronoi tessellation \corr{to obtain} the density
estimate\corr{s $\hat{\rho}_i$ used in} the particle Lagrangian (\ref{eq:L}).
\corr{The tessellation constructs a partition of unity, such that different
  particle distributions with uniform $\hat{\rho}_i$ have \emph{exactly} the
  same average $\hat{\rho}$, i.e.\ the global minimum of $\hat{U}$ is highly
  degenerate. This method has neither a pairing instability, nor `\corr{E$_0$}
  errors', nor the re-ordering capacity of SPH,} but requires additional terms
for that \corr{latter} purpose.

\begin{figure}
  \begin{center}
    \resizebox{72mm}{!}{\includegraphics{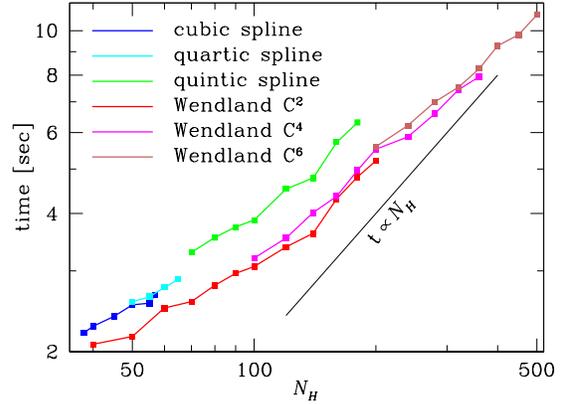}}
  \end{center}
  \vspace*{-3ex}
  \caption{
    \label{fig:time}
    Wall-clock timings for an (average) single SPH time step (using four
    processors) for $N=2^{20}$ particles as function of neighbour number
    $\NH$ for the kernels of Table~\ref{tab:kernel}.}
\end{figure}
\subsection{Are there more useful kernel functions?} \label{sec:disc:remark}
Neither the B-splines nor the Wendland functions have been designed with SPH or
the task of density estimation in mind, but derive from interpolation of
function values $y_i$ for given points $\B{x}_i$.

The B-splines were constructed to exactly interpolate polynomials on a regular
1D grid. However, this for itself is not a desirable property in the context of
SPH, in particular for 2D and 3D.

The Wendland functions were designed for interpolation of scattered
multi-dimensional data, viz
\[ \textstyle
s(\B{x}) = \sum_j \alpha_{\!j}\, w(|\B{x}-\B{x}_{\!j}|).
\]
The coefficients $\alpha_{\!j}$ are determined by matching the interpolant
$s(\B{x})$ to the function values, resulting in the linear equations
\[ \textstyle
y_i = \sum_j \alpha_{\!j} \, w(|\B{x}_i-\B{x}_{\!j}|),
\quad i=1\dots n.
\]
If the matrix $\mathsf{W}_{\!i\!j}=w(|\B{x}_i-\B{x}_{\!j}|)$ is positive
definite for \emph{any} choice of $n$ points $\B{x}_i$, then this equation can
always be solved. Moreover, if the function $w(r)$ has compact support, then
$\B{\mathsf{W}}$ is sparse, which greatly reduces the complexity of the
problem. The Wendland functions were designed to fit this bill.  As a side
effect they have non-negative Fourier transform \citep[according
  to][]{Bochner1933}, which together with their compact support, smoothness, and
computational simplicity makes them ideal for SPH with large $\NH$.

So far, the Wendland functions are the only kernels which are stable against
pairing for all $\NH$ and satisfy all other desirable properties from the list
on page~\pageref{list:kernel}.

\subsection{What is the SPH resolution scale?} \label{sec:disc:resolve}
In smooth flows, i.e.\ in the absence of particle disorder, the only error of
the SPH estimates is the bias induced by the smoothing operation. For example,
assuming a smooth density field
\begin{equation} \label{eq:rho:bias}
  \hat{\rho}_i \approx \rho(\B{x}_i) +
  \tfrac{1}{2}\sigma^2\,\B{\nabla}^2\!\rho(\B{x}_i) + \mathcal{O}(h^4)
\end{equation}
\citep[e.g.][]{Monaghan1985, Silverman1986} with $\sigma$ defined in equation
(\ref{eq:sigma}). Since $\sigma$ also sets the resolution of sound waves
(\S\ref{sec:stable:long}), our definition (\ref{eq:h}), $h=2\sigma$, of the SPH
resolution scale is appropriate for smooth flows. The result (\ref
{eq:rho:bias}) is the basis for the traditional claim of $\mathcal{O}(h^2)$
convergence for smooth flows. True flow discontinuities are smeared out over a
length scale comparable to $h$ (though we have not made a detailed investigation
of this).

\corr{In practice, however, particle disorder affects the performance and, as
  our test simulations demonstrated, the actual resolution of SPH can be much
  worse than the smooth-flow limit suggests.}

\begin{figure}
  \begin{center}
    \resizebox{75mm}{!}{\includegraphics{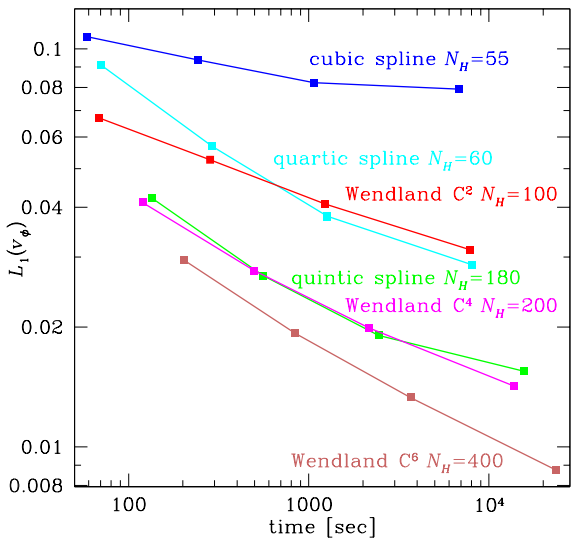}}
  \end{center}
  \vspace*{-3ex}
  \caption{
    \label{fig:gresho:C}
    As Fig.~\ref{fig:gresho}, except that the $x$-axis shows the computational
    costs.}
\end{figure}
\subsection{Are large neighbour numbers sensible?}
There is no consensus about the best neighbour number in SPH: traditionally the
cubic spline kernel is used with $\NH\approx42$, while \cite{Price2012} favours
$\NH=57$ (at or even beyond the pairing-instability limit) and
\cite{ReadHayfieldAgertz2010} use their HOCT4 kernel with even $\NH=442$
(corresponding to a $1.7$ times larger $h$). From a pragmatic point of view, the
number $N$ of particles, the neighbour number $\NH$, and the smoothing kernel
(and between them the numerical resolution) are \emph{numerical parameters}
which can be chosen to optimise the efficiency of the simulation. The critical
question therefore is:
\begin{center}
  \begin{minipage}{73mm}
    \noindent\textbf{\boldmath Which combination of $N$ and $\NH$ (and kernel)
      most efficiently models a given problem at a desired fidelity?  }
  \end{minipage}
\end{center}
Clearly, this will depend on the problem at hand as well as the desired
fidelity. However, if the problem contains any chaotic or turbulent flows, as is
common in star- and galaxy formation, then the situation exemplified in the
Gresho-Chan vortex test of \S\ref{sec:test:vortex} is not atypical and large
$\NH$ may be required for sufficient accuracy.

But are high neighbour numbers affordable? In Fig.~\ref{fig:time}, we plot the
computational cost versus $\NH$ for different kernels. At $\NH\lesssim400$ the
costs rise sub-linearly with $\NH$ (because at low $\NH$ SPH is data- rather
than com\-pu\-tation-dominated) and high $\NH$ are well affordable. In the case
of the vortex test, they are absolutely necessary as Fig.~\ref{fig:gresho:C}
demonstrates: for a given numerical accuracy, our highest $\NH$ makes optimal
use of the computational resources (in our code memory usage does not
significantly depend on $\NH$, so CPU time is the only relevant resource).

\section{Summary} \label{sec:conc}

Particle disorder is unavoidable in strong shear (ubiquitous in astrophysical
flows) and causes random errors of the SPH \correct{force} estimator. The good
news is that particle disorder is less severe than Poissonian shot noise and
the resulting \corr{force} errors \corr{\citep[which are dominated by the
    `E$_0$' term of][]{ReadHayfieldAgertz2010}} are not catastrophic. The bad
news, however, is that these errors are still significant enough to spoil the
convergence of SPH.

In this study we investigated the option to reduce \corr{the `\corr{E$_0$}
  errors'} by increasing the neighbour number in conjunction with a change of
the smoothing kernel. Switching from the cubic to the quintic spline at fixed
resolution $h$ increases the neighbour number $\NH$ only by a
factor\footnote{Using our definition (\ref{eq:h}) for the smoothing scale
  $h$. The conventional factor is 3.375, almost twice 1.74, but \corr{formally}
  effects to a loss of resolution, since the convention\corr{al value} for $h$
  of the B-spline kernels is inappropriate.}  1.74, hardly enough to combat
`\corr{E$_0$} errors'. For a significant reduction of the these errors one has
to trade resolution and significantly increase $\NH$ beyond conventional values.

The main obstacle with this approach is the pairing instability, which occurs
for large $\NH$ with the traditional SPH smoothing kernels. In
\S\ref{sec:linear} and appendix~\ref{app:linear}, we have performed (it appears
for the first time) a complete linear stability analysis for conservative SPH in
any number of spatial dimensions. This analysis shows that SPH smoothing kernels
whose Fourier transform is negative for some wave vector $\B{k}$ will inevitably
trigger the SPH pairing instability at sufficiently large neighbour number
$\NH$. Such kernels therefore require $\NH$ to not exceed a certain threshold in
order to avoid the pairing instability (not to be confused with the tensile
instability, which has the same symptoms but is caused by a negative effective
pressure independent of the kernel properties).

Intuitively, the pairing instability can be understood in terms of the SPH
density estimator: if a paired particle distribution obtains a lower average
estimated density, its estimated total thermal energy $\hat{U}$ is smaller and
hence favourable. Otherwise, the smallest $\hat{U}$ occurs for a regular
distribution, driving the automatic maintenance of particle order, a fundamental
ingredient of SPH.

The \cite{Wendland1995} functions, presented in \S\ref{sec:kern:wend}, have been
constructed, albeit for different reasons, to possess a non-negative Fourier
transform, and be of compact support with simple functional form. The first
property and the findings from our tests in \S\ref{sec:test:noise}
demonstrate the remarkable fact that these kernels are stable against pairing
for \emph{all} neighbour numbers (this disproves the long-cultivated myth that
the pairing instability was caused by a maximum in the kernel gradient). Our 3D
test simulations show that the cubic, quartic, and quintic spline kernels become
unstable to pairing for $\NH>55$, 67, and 190, respectively (see
Fig.~\ref{fig:noise:qmin}), but operating close to these thresholds cannot be
recommended.

A drawback of the Wendland kernels is a comparably large density error at low
$\NH$. As we argue in \S\corr{\ref{sec:pairing:alt}}, this error is directly
related to the stability against pairing. However, in \S\ref{sec:kern:dens} we
present a simple method to correct for this error without affecting the
stability properties and without any other adverse effects.

We conclude, therefore, that the Wendland functions are ideal candidates for SPH
smoothing kernels, in particular when large $\NH$ are desired, since they are
computationally superior to the high-order B-splines. All other alternative
kernels proposed in the literature are computationally more demanding and are
either centrally spiked, like the HOCT4 kernel of \cite{ReadHayfieldAgertz2010},
or susceptible to pairing like the B-splines
\citep[e.g.][]{CabezonGarciaSenzRelano2008}.

Our tests of Section~\ref{sec:test} show that simulations of both strong shear
flows and shocks benefit from large $\NH$. These tests suggest that for
$\NH\sim200$ and 400, respectively, the Wendland $C^4$ and $C^6$ kernels are
most suitable. Compared to $\NH=55$ with the standard cubic spline kernel, these
kernel-$\NH$ combinations have lower resolution ($h$ increased by factors 1.27
and 1.44, respectively), but obtain much better convergence in our tests.

For small neighbour numbers, however, these tests and our linear stability
analysis unexpectedly show that the quartic B-spline kernel with $\NH=60$ is
clearly superior to the traditional cubic spline and can compete with the
Wendland $C^2$ kernel with $\NH=100$. The reason for this astonishing
performance of the quartic spline is \corr{unclear}, perhaps the fact that near
$\B{x}=0$ this spline is more than three times continuously differentiable plays
a role.

We note that, while the higher-degree Wendland functions are new to SPH, the
Wendland $C^2$ kernel has already been used (\citealt{Monaghan2011}, for
example, employs it for 2D simulations). However, while its immunity to the
pairing instability has been noted \citep[e.g.][]{Robinson2009}\footnote{After
  submission of this study, we learned of \citeauthor{Robinson2009}'s
  (\citeyear{Robinson2009}) PhD thesis, where in chapter~7 he compares the
  stability properties of the cubic spline and Wendland $C^2$ kernels in the
  context of 2D SPH simulations. Robinson refutes experimentally the traditional
  explanation (\'a la \citealt{SwegleHicksAttaway1995}) for the pairing
  instability and notices the empirical connection between the pairing
  instability and the non-negativity of the kernel Fourier transform, both in
  excellent agreement with our results.}, we are not aware of any explanation
(previous to ours) nor of any other systematic investigation of the suitability
of the Wendland functions for SPH.

\section*{Acknowledgments}
Research in Theoretical Astrophysics at Leicester is supported by an STFC
rolling grant. We thank Chris Nixon \corr{and Justin Read} for many stimulating
discussions \corr{and} the referee Daniel Price for useful comments \corr{and
  prompt reviewing}.

This research used the ALICE High Performance Computing Facility at the
University of Leicester. Some resources on ALICE form part of the DiRAC Facility
jointly funded by STFC and the Large Facilities Capital Fund of BIS.


\begin{thebibliography}{}

\bibitem[\protect\citeauthoryear{{Abel}}{{Abel}}{2011}]{Abel2011}
{Abel} T.,  2011, \mnras, 413, 271

\bibitem[\protect\citeauthoryear{{Bochner}}{{Bochner}}{1933}]{Bochner1933}
{Bochner} S.,  1933, Math.~Ann., 108, 378

\bibitem[\protect\citeauthoryear{{Cabez{\'o}n}, {Garc{\'{\i}}a-Senz} \&
  {Rela{\~n}o}}{{Cabez{\'o}n} et~al.}{2008}]{CabezonGarciaSenzRelano2008}
{Cabez{\'o}n} R.~M.,  {Garc{\'{\i}}a-Senz} D.,    {Rela{\~n}o} A.,  2008, J.\
  Comp.\ Phys., 227, 8523

\bibitem[\protect\citeauthoryear{{Cullen} \& {Dehnen}}{{Cullen} \&
  {Dehnen}}{2010}]{CullenDehnen2010}
{Cullen} L.,  {Dehnen} W.,  2010, \mnras, 408, 669

\bibitem[\protect\citeauthoryear{{Fulk} \& {Quinn}}{{Fulk} \&
  {Quinn}}{1996}]{FulkQuinn1996}
{Fulk} D.~A.,  {Quinn} D.~W.,  1996, J.\ Comp.\ Phys., 126, 165

\bibitem[\protect\citeauthoryear{{Gingold} \& {Monaghan}}{{Gingold} \&
  {Monaghan}}{1977}]{GingoldMonaghan1977}
{Gingold} R.~A.,  {Monaghan} J.~J.,  1977, \mnras, 181, 375

\bibitem[\protect\citeauthoryear{{Gray}, {Monaghan} \& {Swift}}{{Gray}
  et~al.}{2001}]{GrayMonaghanSwift2001}
{Gray} J.~P.,  {Monaghan} J.~J.,    {Swift} R.~P.,  2001, Comp.\ Meth.\ Appl.\
  Mech.\ Eng., 190, 6641

\bibitem[\protect\citeauthoryear{{Gresho} \& {Chan}}{{Gresho} \&
  {Chan}}{1990}]{GreshoChan1990}
{Gresho} P.~M.,  {Chan} S.~T.,  1990, {Int.\ J.\ Num.\ Meth.\ Fluids}, 11, 621

\bibitem[\protect\citeauthoryear{{Hall}}{{Hall}}{1927}]{Hall1927}
{Hall} P.,  1927, Biometrika, 19, 240

\bibitem[\protect\citeauthoryear{{Herant}}{{Herant}}{1994}]{Herant1994}
{Herant} M.,  1994, \memsai, 65, 1013

\bibitem[\protect\citeauthoryear{{He{\ss}} \& {Springel}}{{He{\ss}} \&
  {Springel}}{2010}]{HessSpringel2010}
{He{\ss}} S.,  {Springel} V.,  2010, \mnras, 406, 2289

\bibitem[\protect\citeauthoryear{{Irwin}}{{Irwin}}{1927}]{Irwin1927}
{Irwin} J.~O.,  1927, Biometrika, 19, 225

\bibitem[\protect\citeauthoryear{{Liska} \& {Wendroff}}{{Liska} \&
  {Wendroff}}{2003}]{LiskaWendroff2003}
{Liska} R.,  {Wendroff} B.,  2003, {SIAM J.\ Sci.\ Comp.}, 25, 995

\bibitem[\protect\citeauthoryear{{Lucy}}{{Lucy}}{1977}]{Lucy1977}
{Lucy} L.~B.,  1977, \aj, 82, 1013

\bibitem[\protect\citeauthoryear{{Monaghan}}{{Monaghan}}{1985}]{Monaghan1985}
{Monaghan} J.~J.,  1985, J.\ Comp.\ Phys., 60, 253

\bibitem[\protect\citeauthoryear{{Monaghan}}{{Monaghan}}{2000}]{Monaghan2000}
{Monaghan} J.~J.,  2000, J.\ Comp.\ Phys., 159, 290

\bibitem[\protect\citeauthoryear{{Monaghan}}{{Monaghan}}{2002}]{Monaghan2002}
{Monaghan} J.~J.,  2002, \mnras, 335, 843

\bibitem[\protect\citeauthoryear{{Monaghan}}{{Monaghan}}{2005}]{Monaghan2005}
{Monaghan} J.~J.,  2005, Reports of Progress in Physics, 68, 1703

\bibitem[\protect\citeauthoryear{{Monaghan}}{{Monaghan}}{2011}]{Monaghan2011}
{Monaghan} J.~J.,  2011, European J.\ Mech.\ B Fluids, 30, 360

\bibitem[\protect\citeauthoryear{{Monaghan}}{{Monaghan}}{2012}]{Monaghan2012}
{Monaghan} J.~J.,  2012, Ann.\ Rev.\ Fluid Mech., 44, 323

\bibitem[\protect\citeauthoryear{{Monaghan} \& {Lattanzio}}{{Monaghan} \&
  {Lattanzio}}{1985}]{MonaghanLattanzio1985}
{Monaghan} J.~J.,  {Lattanzio} J.~C.,  1985, \aap, 149, 135

\bibitem[\protect\citeauthoryear{{Monaghan} \& {Price}}{{Monaghan} \&
  {Price}}{2001}]{MonaghanPrice2001}
{Monaghan} J.~J.,  {Price} D.~J.,  2001, \mnras, 328, 381

\bibitem[\protect\citeauthoryear{{Morris}}{{Morris}}{1996}]{Morris1996}
{Morris} J.~P.,  1996, Pub.\ Astr.\ Soc.\ Aust., 13, 97

\bibitem[\protect\citeauthoryear{{Nelson} \& {Papaloizou}}{{Nelson} \&
  {Papaloizou}}{1994}]{NelsonPapaloizou1994}
{Nelson} R.~P.,  {Papaloizou} J.~C.~B.,  1994, \mnras, 270, 1

\bibitem[\protect\citeauthoryear{{Phillips} \& {Monaghan}}{{Phillips} \&
  {Monaghan}}{1985}]{PhillipsMonaghan1985}
{Phillips} G.~J.,  {Monaghan} J.~J.,  1985, \mnras, 216, 883

\bibitem[\protect\citeauthoryear{{Price}}{{Price}}{2012}]{Price2012}
{Price} D.~J.,  2012, J.\ Comp.\ Phys., 231, 759

\bibitem[\protect\citeauthoryear{{Rasio}}{{Rasio}}{2000}]{Rasio2000}
{Rasio} F.~A.,  2000, Prog.\ Theo.\ Phys.\ Supp., 138, 609

\bibitem[\protect\citeauthoryear{{Read} \& {Hayfield}}{{Read} \&
  {Hayfield}}{2012}]{ReadHayfield2012}
{Read} J.~I.,  {Hayfield} T.,  2012, \mnras, in press

\bibitem[\protect\citeauthoryear{{Read}, {Hayfield} \& {Agertz}}{{Read}
  et~al.}{2010}]{ReadHayfieldAgertz2010}
{Read} J.~I.,  {Hayfield} T.,    {Agertz} O.,  2010, \mnras, 405, 1513

\bibitem[\protect\citeauthoryear{{Robinson}}{{Robinson}}{2009}]{Robinson2009}
{Robinson} M.~J.,  2009, PhD thesis, Monash University, Australia

\bibitem[\protect\citeauthoryear{{Rosswog}}{{Rosswog}}{2009}]{Rosswog2009}
{Rosswog} S.,  2009, New Astronomy Reviews, 53, 78

\bibitem[\protect\citeauthoryear{{Schoenberg}}{{Schoenberg}}{1946}]{Schoenberg%
1946}
{Schoenberg} I.~J.,  1946, Q.~Appl.~Math., 4, 45

\bibitem[\protect\citeauthoryear{{Sch{\"u}{\ss}ler} \&
  {Schmitt}}{{Sch{\"u}{\ss}ler} \& {Schmitt}}{1981}]{SchuesslerSchmitt1981}
{Sch{\"u}{\ss}ler} I.,  {Schmitt} D.,  1981, \aap, 97, 373

\bibitem[\protect\citeauthoryear{{Silverman}}{{Silverman}}{1986}]{Silverman198%
6}
{Silverman} B.~W.,  1986, {Density estimation for statistics and data
  analysis}.
London, Chapman \& Hall

\bibitem[\protect\citeauthoryear{{Sod}}{{Sod}}{1978}]{Sod1978}
{Sod} G.~A.,  1978, J.\ Comp.\ Phys., 27, 1

\bibitem[\protect\citeauthoryear{{Springel}}{{Springel}}{2010}]{Springel2010:r%
eview}
{Springel} V.,  2010, \araa, 48, 391

\bibitem[\protect\citeauthoryear{{Springel} \& {Hernquist}}{{Springel} \&
  {Hernquist}}{2002}]{SpringelHernquist2002}
{Springel} V.,  {Hernquist} L.,  2002, \mnras, 333, 649

\bibitem[\protect\citeauthoryear{{Swegle}, {Hicks} \& {Attaway}}{{Swegle}
  et~al.}{1995}]{SwegleHicksAttaway1995}
{Swegle} J.~W.,  {Hicks} D.~L.,    {Attaway} S.~W.,  1995, J.\ Comp.\ Phys.,
  116, 123

\bibitem[\protect\citeauthoryear{{Thomas} \& {Couchman}}{{Thomas} \&
  {Couchman}}{1992}]{ThomasCouchman1992}
{Thomas} P.~A.,  {Couchman} H.~M.~P.,  1992, \mnras, 257, 11

\bibitem[\protect\citeauthoryear{{Wendland}}{{Wendland}}{1995}]{Wendland1995}
{Wendland} H.,  1995, {Adv.~Comp.~Math.}, 4, 389

\bibitem[\protect\citeauthoryear{{Wendland}}{{Wendland}}{2005}]{Wendland2005}
{Wendland} H.,  2005, {Scattered Data Approximation}.
Cambridge University Press, Cambridge, UK

\end{thebibliography}

\appendix
\section{Linear Stability Analysis} \label{app:linear}
We start from an equilibrium with particles of equal mass $m$ on a regular
grid and impose a plane-wave perturbation to the unperturbed positions
$\bar{\B{x}}_{i}$ (a bar denotes a quantity obtained for the unperturbed
equilibrium):
\begin{equation} \label{eq:x:pert}
  \B{x}_{i}^{} = \bar{\B{x}}_{i}^{} + \B{\xi}_i,
  \qquad
  \B{\xi}_i = \B{a}\,\Phi_i(\B{x},t),
  \qquad
  \Phi_i(\B{x},t) = \mathrm{e}^{i(\B{k}{\cdot}\bar{\B{x}}_{i}-\omega t)}.
\end{equation}
as in equation (\ref{eq:x:pert:}). We derive the dispersion relation
$\omega(\B{k},\B{a})$ by equating the SPH force imposed by the perturbation
(to linear order) to its acceleration
\begin{equation} \label{eq:ddxi:X}
  \ddot{\B{\xi}}_i=
  -\B{a}\,\omega^2\,\Phi_i(\B{x},t).
\end{equation}
To obtain the perturbed SPH forces to linear order, we develop the internal
energy of the system, and hence the SPH density estimate, to second order in
$\B{\xi}_i$. If $\hat{\rho}_i=\bar{\rho}+\rho_{i1} +\rho_{i2}$ with
$\rho_{i1}$ and $\rho_{i2}$ the first and second-order density corrections,
respectively, then
\begin{equation}
  \label{eq:hydro:P}
  \ddot{\B{\xi}}_i = -\frac{1}{m}\frac{\partial\mathcal{L}}{\partial\B{\xi}_i}
  = 
  -
  \frac{\bar{P}}{\bar{\rho}}
  \sum_k
  \frac{1}{\bar{\rho}}
  \frac{\partial\rho_{k2}}{\partial\B{\xi}_i}
  -
  \left(
    \bar{c}^2-\frac{2\bar{P}}{\bar{\rho}}
  \right)
  \frac{1}{2\bar{\rho}^2}
  \sum_k 
  \frac{\partial\rho_{k1}^2}{\partial\B{\xi}_i}.
\end{equation}
\subsection{Fixed \boldmath $h$}
Let us first consider the simple case of constant $h=\bar{h}$ which
remains unchanged during the perturbation. Then,
\begin{equation} \label{eq:varrho}
  \textstyle
  \rho_{in} = \varrho_{in} / n!
  \quad\text{with}\quad
  \varrho_{in} = \sum_j m \,(\B{\xi}_{i\!j}{\cdot}\B{\nabla})^n
  W(\bar{\B{x}}_{i\!j},\bar{h}).
\end{equation}
Inserting
\begin{equation} \label{eq:xiij}
  \B{\xi}_{i\!j} = \B{\xi}_i-\B{\xi}_{\!j} = \B{a}
  \big(1-\mathrm{e}^{-i\B{k}{\cdot}\bar{\B{x}}_{i\!j}}\big)\, \Phi_i
\end{equation}
into (\ref{eq:varrho}) gives
\begin{equation} 
  \textstyle
  \varrho_{i1} = \Phi_i\,
  \sum_jm\,\big(1-\mathrm{e}^{-i\B{k}{\cdot}\bar{\B{x}}_{i\!j}}\big)\,
  \B{a}{\cdot}\B{\nabla}W(\bar{\B{x}}_{i\!j},\bar{h})
  = i\, \Phi_i\,\B{a}{\cdot}\B{t},
\end{equation}
where (assuming a symmetric particle distribution)
\begin{equation} \label{eq:t}
  \textstyle 
  \B{t}(\B{k}) = 
  \sum_j m \sin\B{k}{\cdot}\bar{\B{x}}_{\!j}\,\B{\nabla}
  W(\bar{\B{x}}_{\!j},\bar{h}).
\end{equation}
We can then derive
\begin{eqnarray}
  \textstyle
  \sum_k \varrho_{k1}\, \partial\varrho_{k1}/\partial\B{\xi}_i
  &=& \textstyle
  \sum_{k}\varrho_{k1}\, \sum_j m\,\B{\nabla}W(\bar{\B{x}}_{k\!j},\bar{h})
  \big(\delta_{ik}-\delta_{i\!j}\big)
  \nonumber \\
  &=& \textstyle
  \sum_j m\,(\varrho_{i1}+\varrho_{j1})\B{\nabla}W(\bar{\B{x}}_{i\!j},\bar{h})
  \nonumber \\
  &=& \textstyle
  i\,\Phi_i\,\B{a}{\cdot}\B{t}
  \sum_jm\,\big(1+\mathrm{e}^{-i\B{k}{\cdot}\bar{\B{x}}_{i\!j}}\big)\,
  \B{\nabla}W(\bar{\B{x}}_{i\!j},\bar{h})
  \nonumber \\ \label{eqs:derive:1}
  &=&
  \Phi_i\,\B{a}{\cdot}\B{t}\,\B{t},
  \\
  \textstyle
  \tfrac{1}{2}\sum_{k}\partial\varrho_{k2}/\partial\B{\xi}_{i}
  &=& \textstyle 
  \sum_{k\!j} m \,(\B{\xi}_{k\!j}{\cdot}\B{\nabla})\B{\nabla}
  W(\bar{\B{x}}_{k\!j},\bar{h})\big(\delta_{ik}-\delta_{i\!j}\big)
  \nonumber \\
  &=& \textstyle
  2\sum_j m \,(\B{\xi}_{i\!j}{\cdot}\B{\nabla})\B{\nabla}
  W(\bar{\B{x}}_{i\!j},\bar{h})
  \nonumber \\
  &=& \textstyle
  2\Phi_i\,\sum_j m\,
  \big(1-\mathrm{e}^{-i\B{k}{\cdot}\bar{\B{x}}_{i\!j}}\big)\,
  \B{a}{\cdot}\B{\nabla}\B{\nabla}W(\bar{\B{x}}_{k\!j},\bar{h})
  \nonumber \\ \label{eqs:derive:2}
  &=& 
  2\Phi_i\,\B{a}{\cdot}\B{\mathsf{T}}
\end{eqnarray}
with
\begin{equation} \label{eq:T}
  \textstyle 
  \B{\mathsf{T}}(\B{k}) =
  \sum_j m \,\big(1-\cos\B{k}{\cdot}\bar{\B{x}}_{\!j}\big)\,
  \B{\nabla}^{(2)}W(\bar{\B{x}}_{\!j},\bar{h}).
\end{equation}
Inserting these results into (\ref{eq:hydro:P}), we get
\begin{eqnarray}
  \ddot{\B{\xi}}_i &=& -
  \frac{2\bar{P}}{\bar{\rho}}\frac{\B{a}{\cdot}\B{\mathsf{T}}}{\bar{\rho}}\Phi_i
  - \left(\bar{c}^2-\frac{2\bar{P}}{\bar{\rho}}\right)
  \frac{\B{a}{\cdot}\B{t}\,\B{t}}{\bar{\rho}^2} \Phi_i.
\end{eqnarray}

\subsection{Adaptive smoothing}
If the $h_i$ are adapted such that $M_i\equiv h_i^\nu \hat{\rho}_i$ remains a
global constant $\bar{M}$, the estimated density is simply
$\hat{\rho}_i=\bar{M}h_i^{-\nu}$. We start by expanding $M_i$ to second order
in both $\B{a}$ and $\eta_i\equiv\ln(h_i/\bar{h})$. Using a prime to denote
differentiation w.r.t.\ $\ln\bar{h}$, we have
\begin{subequations}
  \begin{eqnarray}
    M_i &=& M_h
    +\bar{h}^\nu\!\varrho_{i1}
    +\eta\big(\bar{h}^\nu\bar{\rho}\big)'
    +\tfrac{1}{2}\bar{h}^\nu\!\varrho_{i2}
    +\eta\big(\bar{h}^\nu\!\varrho_{i1}\big)'
    +\tfrac{1}{2}\eta^2 \big(\bar{h}^\nu\bar{\rho}\big)''
    \\ &=&
    M_h + \bar{h}^\nu\left[
      \varrho_{i1}
      +\eta\nu\bar{\rho}\bar{\Omega}
      +\tfrac{1}{2}\varrho_{i2}
      +\eta\varrho_{i1}'
      +\eta\nu\varrho_{i1}
      +\tfrac{1}{2}\eta^2\nu^2\bar{\rho}\bar{\Pi}
      \right]
  \end{eqnarray}
\end{subequations}
with $\bar{\Omega}\simeq1$ as defined in
equation (\ref{eq:Omega}) and
\begin{equation}
  \label{eq:Pi}
  \bar{\Pi} =
  \frac{1}{\nu^2\bar{h}^\nu\bar{\rho}}\,
  \frac{\partial^2(\bar{h}^\nu\bar{\rho})}{\partial(\ln\bar{h})^2}
  = \frac{1}{\nu\bar{h}^\nu\bar{\rho}}\,
  \frac{\partial(\bar{h}^\nu\bar{\rho}\bar{\Omega})}{\partial\ln\bar{h}}
  \simeq 1.
\end{equation}
By demanding $M_i=\bar{M}$, we obtain for the first- and second-order
contributions to $\eta_i^{}$
\begin{subequations}
  \label{eq:eta:12}
  \begin{eqnarray}
    \label{eq:eta:1}
    -\nu\bar{\rho}\eta_{i1}^{} &=&
    \frac{\varrho_{i1}}{\bar{\Omega}}
    \\ \label{eq:eta:2p}
    -\nu\bar{\rho}\eta_{i2}^{} &=&
    \frac{\varrho_{i2}}{2\bar{\Omega}}
    +\frac{\eta_{i1}\varrho_{i1}'}{\bar{\Omega}}
    +\frac{\nu\eta_{i1}\varrho_{i1}}{\bar{\Omega}}
    +\frac{\nu^2\eta_{i1}^2\bar{\rho}\bar{\Pi}}{2\bar{\Omega}}
    \\ \label{eq:eta:2}
    &=&
    \frac{\varrho_{i2}}{2\bar{\Omega}}
    -\frac{(\varrho_{i1}^2)'}{2\nu\bar{\rho}\bar{\Omega}^2}
    -\frac{\varrho_{i1}^2}{\bar{\rho}\bar{\Omega}^2}
    +\frac{\varrho_{i1}^2\bar{\Pi}}{2\bar{\rho}\bar{\Omega}^3}.
  \end{eqnarray}
\end{subequations}
From these expressions and $\hat{\rho}_i=\bar\rho\mathrm{e}^{-\nu\eta_i}$ we
obtain the first and second order density corrections
\begin{subequations}
  \label{eq:rho:12:A}
  \begin{eqnarray}
    \label{eq:rho:1:A}
    \rho_{i1}^{} &= 
    -\nu\bar{\rho}\eta_{i1}
    \phantom{\;\,+ \tfrac{1}{2}\nu^2\bar{\rho} \eta_{i1}^2} =&
    \frac{\varrho_{i1}}{\bar{\Omega}},
    \\
    \label{eq:rho:2:A}
    \rho_{i2}^{} &= 
    -\nu\bar{\rho}\eta_{i2} + \tfrac{1}{2}\nu^2\bar{\rho} \eta_{i1}^2 =&
    \frac{\varrho_{i2}}{2\bar{\Omega}}
    -\frac{(\varrho_{i1}^2)'}{2\nu\bar{\rho}\bar{\Omega}^2}
    +\frac{\varrho_{i1}^2(\bar{\Pi}-\bar{\Omega})}
            {2\bar{\rho}\bar{\Omega}^3}.
  \end{eqnarray}
\end{subequations}
Inserting these expressions into equation (\ref{eq:hydro:P}) we find with
relations (\ref{eqs:derive:1}) and (\ref{eqs:derive:2})
\begin{eqnarray}
  \ddot{\B{\xi}}_i &=& 
  -\frac{\bar{P}}{\bar{\rho}}
  \left(
  2\frac{\B{a}{\cdot}\B{\mathsf{T}}}{\bar{\rho}\bar{\Omega}}
  + \bar{\Xi}\frac{\B{a}{\cdot}\B{t}\,\B{t}}{\bar{\rho}^2\bar{\Omega}^2}
  - \frac{1}{\nu\bar{\rho}^2\bar{\Omega}^2}
  \frac{\partial(\B{a}{\cdot}\B{t}\,\B{t})}{\partial\ln\bar{h}}
  \right)\Phi_i
  \nonumber \\[0.5ex] && \label{eq:ddxi:ad}
  -
  \left( \bar{c}^2-\frac{2\bar{P}}{\bar{\rho}} \right)
  \frac{\B{a}{\cdot}\B{t}\,\B{t}}{\bar{\rho}^2\bar{\Omega}^2}\,\Phi_i
  .
\end{eqnarray}
where
\begin{equation}
  \bar{\Xi} \equiv \frac{\bar{\Pi}}{\bar{\Omega}}-1
  = \frac{1}{\nu}
  \frac{\partial\ln(\bar{\rho}\bar{\Omega})}{\partial\ln\bar{h}} \simeq 0.
\end{equation}
\subsection{The limit \boldmath $\B{k}\to0$} \label{app:limit}
From equation (\ref{eq:W:h})
$h^\nu\B{x}{\cdot}\B{\nabla}W(\B{x},h)=-\partial(h^\nu W)/\partial\ln h$,
such that 
\begin{subequations}
  \begin{eqnarray}
    -\nu\,\hat{\rho}_i\Omega_i &=& \textstyle
    \sum_j m_j \bxij{\cdot}\B{\nabla}W(\bxij,h_i),
    \\
    \nu^2\hat{\rho}_i\Pi_i &=& \textstyle
    \sum_j m_j (\bxij{\cdot}\B{\nabla})^2W(\bxij,h_i).
  \end{eqnarray}
\end{subequations}
Assuming local spatial isotropy we then get in the limit $\B{k}\to0$
\begin{subequations}
  \begin{eqnarray}
    \B{t} &\to& - \bar{\Omega}\bar{\rho}\,\B{k},
    \\
    \B{t}' &\to& - \nu(\bar{\Pi}-\bar{\Omega})\bar{\rho}\,\B{k},
    \\
    \B{\mathsf{T}} &\to& \frac{(2\bar{\Omega}+\nu\bar{\Pi})\bar{\rho}}
      {2+\nu}\B{k}^{(2)} 
    + \frac{\nu(\bar{\Pi}-\bar{\Omega})\bar{\rho}}{2(\nu+2)}|\B{k}|^2
    \,\B{\mathsf{I}}.
  \end{eqnarray}
\end{subequations}
Inserting these into (\ref{eq:ddxi:ad}) gives
\begin{equation}
  \ddot{\B{\xi}}=-\bar{c}^2 \B{a}{\cdot}\B{k}\,\B{k}\,\Phi_i
  -\frac{\bar{P}}{\bar{\rho}}
  \bar{\Xi}
  \left[\bigg(\!\frac{2\nu}{2+\nu}-1\!\bigg)
  \,\B{a}{\cdot}\B{k}\,\B{k}
  +\frac{\nu\B{k}^2\B{a}}{\nu+2}
  \right]\Phi_i.
  \label{lastpage}
\end{equation}

\end{document}